\documentclass[11pt,a4paper]{article}

\usepackage[T1]{fontenc}
\usepackage{graphicx}
\usepackage[subrefformat=parens]{subcaption}
\usepackage[dvipsnames]{xcolor}
\usepackage{amsmath}
\usepackage{amssymb}

\usepackage[sorting=none,giveninits=true]{biblatex}
\bibliography{bibfile.bib}

\usepackage{tabularx} 
\usepackage{booktabs}
\usepackage{array} 
\usepackage{multirow}
\usepackage{textcomp,gensymb}
\usepackage{siunitx}
\usepackage{isotope}
\usepackage{hyperref}
\usepackage{authblk}
\usepackage[normalem]{ulem} 
\hypersetup{
    colorlinks=true,
    linkcolor=blue,
    citecolor=magenta,
    }
\usepackage{tikz}

\usepackage{fullpage}

\graphicspath{{fig/}}

\newlength{\stdfigwidth}
\newcommand{\geant}{{\textsc{Geant4}}}
\newcommand{\fluka}{{\textsc{Fluka}}}
\newcommand{\polyethylene}{\ensuremath{\text{C}_{n}\text{H}_{2n}}}
\newcommand{\mwe}{\text{m\,w.\,e.}}
\newcommand{\mwee}{\text{m\,w.\,e}}
\newcommand{\nfluxunit}{\si{\ensuremath{n/\mu}/(\gram/\cm\squared)}}

\title{Muon-induced background in a next-generation \\ dark matter experiment based on liquid xenon}

\begin{document}

\author[1,3,$\dagger$]{Viktor P\v{e}\v{c}}
\author[1]{Vitaly A. Kudryavtsev}
\author[2]{Henrique M. Ara\'ujo}
\author[2]{Timothy J. Sumner}

\affil[1]{Department of Physics and Astronomy, University of Sheffield, Sheffield, S3 7RH, UK}
\affil[2]{Department of Physics, Imperial College London, London, SW7 2AZ, UK}
\affil[3]{FZU -- Institute of Physics of the Czech Academy of Sciences, Prague, 182 00, CZE}

\maketitle
{
\renewcommand{\thefootnote}{\fnsymbol{footnote}}
\footnotetext[2]{corresponding author: viktor.pec@fzu.cz}
}

\begin{abstract}
Muon-induced neutrons can lead to potentially irreducible backgrounds in rare event search experiments. We have investigated the implication of laboratory depth on the muon-induced background in a future dark matter experiment capable of reaching the so-called neutrino floor. Our simulation study focused on a xenon-based detector with 70\,tonnes of active mass, surrounded by additional veto systems plus a water shield. Two locations at the Boulby Underground Laboratory (UK) were analysed as examples: an experimental cavern in salt at a depth of 2850\,\mwe\ (similar to the location of the existing laboratory), and a deeper laboratory located in polyhalite rock at a depth of 3575\,\mwee. Our results show that no cosmogenic background events are likely to survive standard analysis cuts for 10~years of operation at either location. 
The largest background component we identified comes from beta-delayed neutron emission from $^{17}$N which is produced from $^{19}$F in the fluoropolymer components of the experiment. Our results confirm that a dark matter search with sensitivity to the neutrino floor is viable (from the point of view of cosmogenic backgrounds) in underground laboratories at these levels of rock overburden. This work was conducted in 2019--21 in the context of a feasibility study to investigate the possibility of developing the Boulby Underground Laboratory to host a next-generation dark matter experiment; however, our findings are also relevant for other underground laboratories.

\end{abstract}

\section{Introduction}

One of the key factors influencing the choice of underground laboratory for a future, high-sensitivity experiment for rare event searches is the depth (overburden) of the site. Events triggered by cosmic-ray muons can be a background to dark matter and neutrinoless double-beta decay ($0\nu\beta\beta$) searches, as well as to some low-energy neutrino experiments detecting signals from nuclear reactors or astrophysical sources. As an example, isolated neutrons produced in muon interactions or muon-induced cascades, if scattering only once in the detector, will mimic nuclear recoils (NR) caused by Weakly Interacting Massive Particles (WIMPs). Most of the (low-energy) events caused by electromagnetic interaction yielding electron recoils (ER) can be removed by powerful discrimination techniques focusing on differences in ionization and primary scintillation signals caused by electrons and recoiling nuclei. For neutrinoless double-beta decay experiments, the muon-induced neutrons may produce a background via radiative capture processes or inelastic scattering. Activation of detector components mainly by hadrons (including neutrons) in muon-induced cascades can add to the background, in particular if this activation results in subsequent decays on time scales of a few seconds or more. In this case, tagging a background event via the detection of the muon may significantly increase the dead time of the experiment, making such tagging inefficient. 

\begin{figure}[!t]
\centerline{\includegraphics[width=12cm]{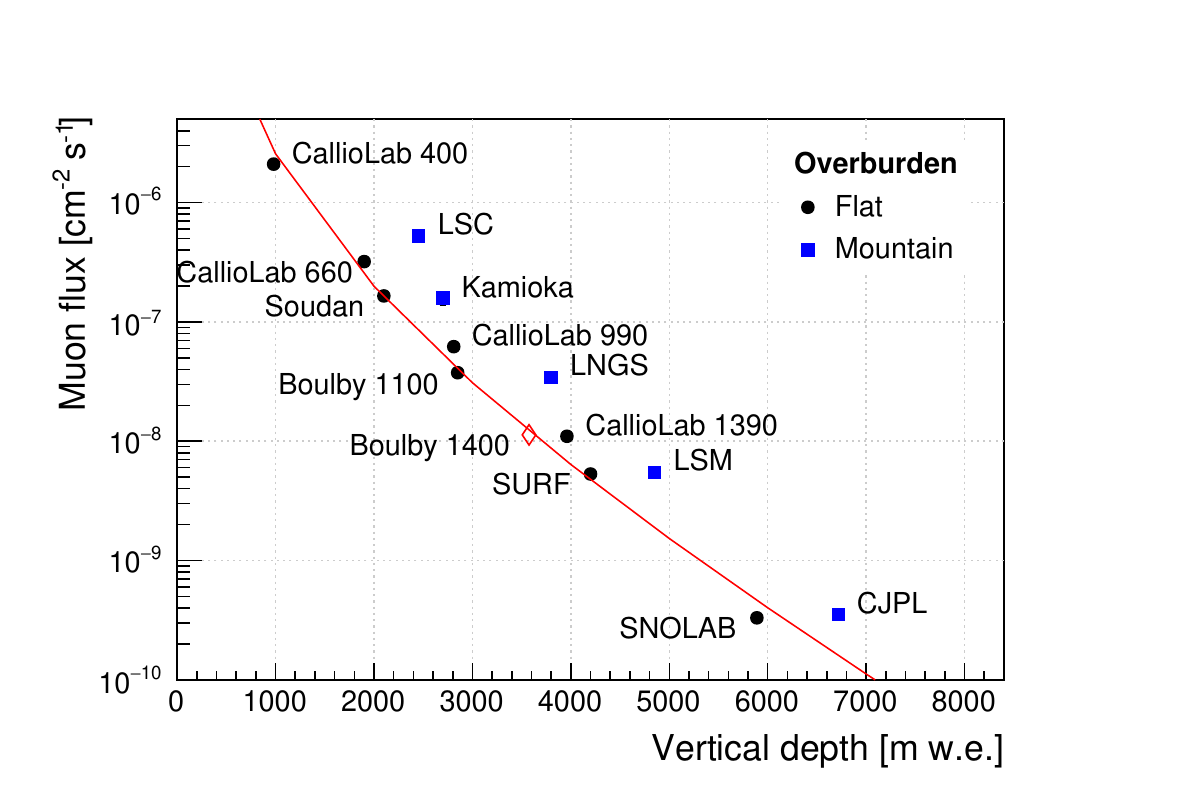}}
\caption[]{Muon flux as a function of vertical depth in metres water equivalent (\mwe) for laboratories around the world. Black markers represent measurements in laboratories with relatively flat surface profile and blue squares represent laboratories under mountains. The red curve is based on Monte Carlo simulations of muons propagating through flat overburden of `standard rock' \cite{music} ($\left<Z\right>=11$, $\left<A\right>=22$). The red open diamond represents the estimated flux at a deeper location at Boulby (1400\,m, or 3575\,\mwe). The data points represent the following measurements:
{CallioLab} (Pyh\"{a}salmi, Finland) at various depths \cite{ENQVIST2005286},
{LSC} (Canfranc, Spain) \cite{Trzaska2019} (depth taken from \cite{Canfranc2005}), {Soudan} (MN, USA, no longer operational) \cite{PhysRevD.90.122003},
{Kamioka} (Japan) \cite{PhysRevD.93.012004} (conversion from muon rate to flux based on MC simulations from \cite{PhysRevD.74.053007}),
{Boulby} at 1100\,m level (2850~\mwe) \cite{lr2013}, 
{LNGS} (Gran Sasso, Italy) \cite{Agostini.2019},
{SURF} (Lead, USA) \cite{ABGRALL201770} (depth taken from \cite{PhysRevD.27.1444}),
{LSM} (Modane, France) \cite{PhysRevD.40.2163},
{SNOLab} (Sudbury, Canada) \cite{PhysRevD.80.012001},
{CJPL} (Jingping, China) \cite{Guo.2021}.
Reported errors of the measurements are too small to be visible in the plot.\protect\footnotemark%
}
\label{fig:muon_flux}
\end{figure}

The muon flux drops rapidly with depth, and a minimum depth requirement is one of the main factors that affect the choice of an underground site for an experiment with a specific designed sensitivity. 
Figure \ref{fig:muon_flux} shows the muon fluxes as measured in different underground laboratories as a function of vertical depth in meters water equivalent (\mwe), together with a calculated depth--intensity curve for standard rock ($\left<Z\right>=11$, $\left<A\right>=22$). The deviation of some of the data points from the calculated curve are mainly due to \textit{a}) the complex surface profile at some locations, and \textit{b}) different elemental composition of rock above the laboratory.
\nocite{music,ENQVIST2005286,Trzaska2019,Canfranc2005,PhysRevD.90.122003,PhysRevD.93.012004,PhysRevD.74.053007,lr2013,Agostini.2019,ABGRALL201770,PhysRevD.27.1444,PhysRevD.40.2163,PhysRevD.80.012001,Guo.2021}

\footnotetext{There is an additional uncertainty of about $12-20\%$ in the convention used for the flux reporting. The MC prediction is for the flux through a spherical detector and so are some measurements (Boulby, LNGS, SURF, etc). Some other experiments reported the flux through a horizontal surface. In some references it was not clear what convention was used. The difference between the two fluxes decreases with depth and is about 15\% at 3\,k\mwe}

Several underground laboratories have muon fluxes similar to that measured at the existing Boulby facility at 1100\,m, and a number of deeper sites achieve fluxes up to two orders of magnitude lower. Our study probed whether the depth such as at the Boulby site was sufficient to enable a dark matter search reaching the neutrino floor -- requiring a peak cross-section sensitivity for spin-independent WIMP-nucleon scattering approaching $\sim$10$^{-49}$~cm$^2$. Although other rare event searches were considered, our simulation work focused on a liquid xenon experiment with 70\,tonnes of active mass (a 10-fold upscale of LUX-ZEPLIN (LZ)~\cite{akerib2017}), as this has been the leading technology in the field.

Neutrons are produced by muons underground by five main processes:
\nopagebreak
\begin{enumerate}
    \item Negative muon capture, dominating at shallow sites with a large fraction of stopping muons;
    \item Direct muon-induced nucleus spallation;
    \item Hadroproduction (including neutron multiplication via neutron inelastic scattering) that originates primarily in the hadronic cascades caused by muon inelastic scattering (also called muon nuclear interaction);
    \item Photoproduction that takes place primarily in electromagnetic cascades generated by muons via bremsstrahlung, $e^{+}e^{-}$-pair production and $\delta$-electron emission processes;
    \item Delayed neutrons due to activation of light isotopes by hadrons, followed by a beta decay accompanied by neutron emission; this process can be attributed to hadroproduction but deserves separate consideration due to a non-negligible time delay between the muon and the neutron.
\end{enumerate}

Neutron production by muons depends on the muon energy and on the composition of the material that the muons pass through. Cosmic-ray muons have a broad energy spectrum at any underground site (see, for instance, Ref. \cite{music} for calculated spectra at different depths) and the mean muon energy is a convenient parameter to characterise the spectrum at a particular depth. It has been shown previously (see, for instance, Refs. \cite{wang2001,vk2003,mei2006,ha2005,al2009} and references therein) that the neutron yield is approximately proportional to $E_{\mu}^{0.7-0.8}$, where the mean muon energy $E_{\mu}$ can replace the muon spectrum at a particular site. Given the complexity of physics processes involved in neutron production, this dependence on muon energy is approximate and works only for high-energy muons (above 10\,GeV) where negative muon capture can be neglected on a scale of a few metres (typical detector size).

The dependence of the neutron yield on material composition reflects the contribution of various mechanisms to neutron production. This has been studied in a variety of papers (see, for instance, Refs. \cite{wang2001,vk2003,mei2006,ha2005,al2009} and references therein) and a proportionality of $A^{0.75-0.80}$ has been observed, where $A$ is the mean atomic weight of a material. This is only a trend and, to calculate neutron production accurately, a detailed simulation including an approximate detector geometry and its surroundings needs to be carried out. Note that the contribution of different processes to neutron production changes with increasing $A$, since the probability of electromagnetic cascade production per atom rises approximately as $Z(Z+1)/A$, whereas that for a hadronic cascade slightly decreases as $A^{-0.2}$.

A number of experiments measured muon-induced neutron production rate underground in different targets. Examples include measurements carried out at the Boulby Underground Laboratory (for lead) \cite{ha2008,lr2013}, LNGS (for scintillator and steel) \cite{borexino2013,lvd2013} and Kamioka (for scintillator) \cite{kamland2010}. Based on the difference between the measured and simulated rates of neutrons reported in Ref. \cite{lr2013}, the authors concluded that the overall uncertainty in neutron production rate was about 25\%. Similar conclusions were reported in several other publications.

The variety of neutron production mechanisms and their dependence on muon energy and material composition, adding to the complex geometry of the setup that includes the detector, shielding, veto systems and the cavern geometry, require full Monte Carlo modelling of all particles produced in muon-induced cascades. These simulations should account for the realistic correlations between neutron multiplicities, energies and angles of neutron emission with respect to the parent muon, and correlations with other particles able to deposit energy in the various detector systems. These correlations are exploited in various techniques of background suppression by tagging the primary muon or muon-induced cascade.

Below, we describe a set of simulations carried out to characterise cosmogenic backgrounds in a next-generation experiment with a large liquid xenon target. We made use of the \geant{} simulation toolkit \cite{geant4,Allison:2006ve,geant4-1}. Section~\ref{validation} presents tests of neutron production as simulated in  \geant~version~10.5 and comparisons with previous simulations and data.  Section~\ref{bkgsim} includes the description of the simplified xenon detector geometry and the simulation procedure. Results are reported in Section~\ref{results}. 
We considered, as examples, two possible locations at Boulby: a site near the existing laboratory in the NaCl layer and a potential site, deeper by 300\,m, in a polyhalite layer. We draw some generic conclusions in Section~\ref{conlusions}.

This work was conducted in 2019-2021 as part of the STFC-funded project on the feasibility of Boulby Underground Laboratory to host a future dark matter experiment, however, its findings are also relevant for other underground laboratories. Preliminary results have been reported in~\cite{Pec:2022dco}.

\section{Neutron production in \protect\geant}\label{validation}
%

We began by conducting a comparison of neutron production rates against previous simulations and data in order to validate key physics processes implemented in \geant{}. In the current modelling of muon events, we have used the version 10.5 of the toolkit (using its data libraries nominal for the toolkit version, i.e.~G4NDL4.5 for neutron-related physics below 20\,MeV, G4PhotonEvaporation-5.3, G4RadioactiveDecay-5.3, etc.) with the `Shielding' physics list, a common choice in simulations for low-background experiments.

\begin{figure}[!htb]
    \makebox[\textwidth]{
        \begin{subfigure}{0.49\textwidth}
        \includegraphics[width=1.1\stdfigwidth]{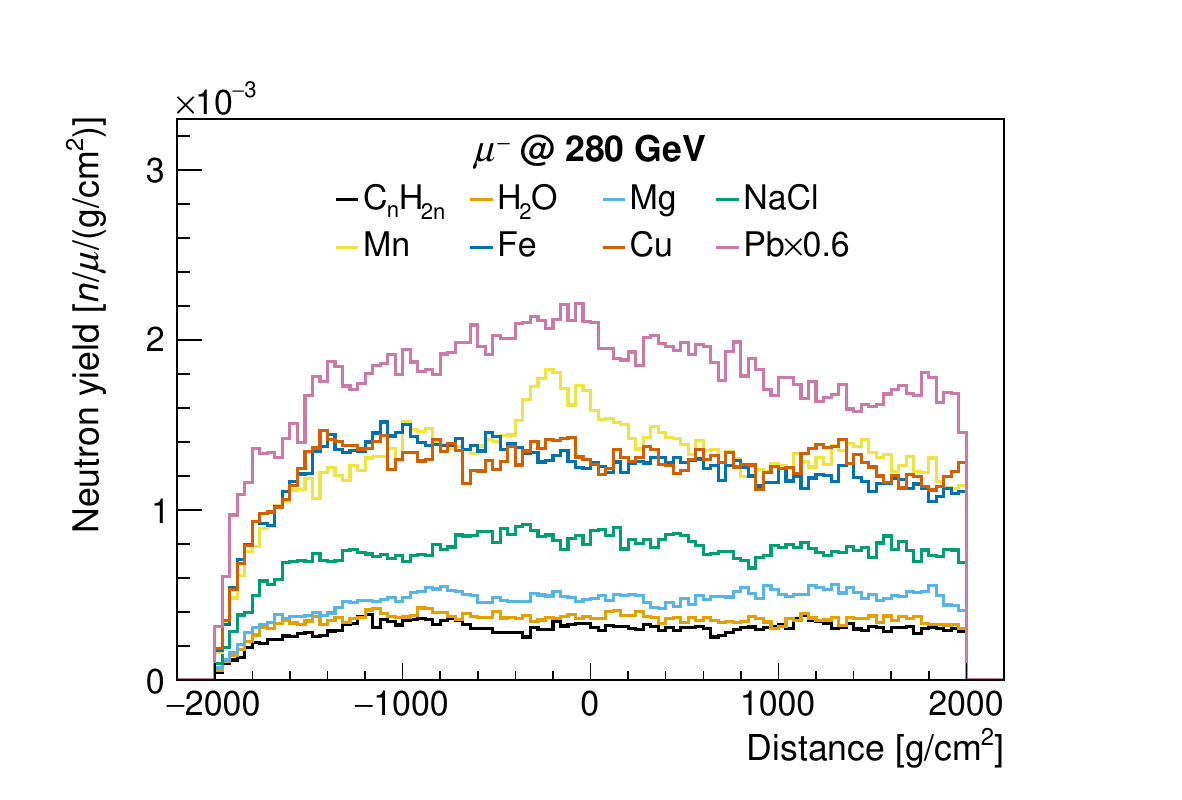}%
        \caption{}\label{fig:n_z:a}
        \end{subfigure}
        \begin{subfigure}{0.49\textwidth}
        \includegraphics[width=1.1\stdfigwidth]{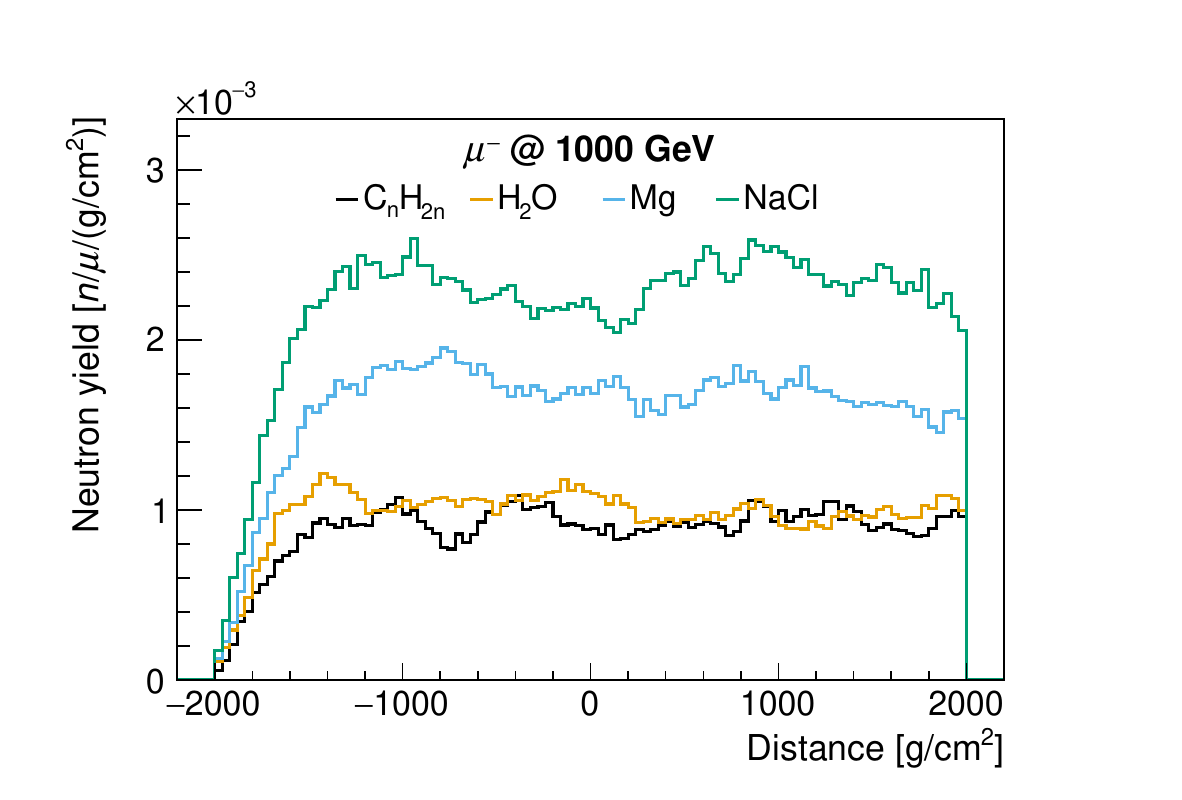}
        \caption{}\label{fig:n_z:b}
        \end{subfigure}
    }
    \caption{Neutron production rate along the muon initial direction. \num{100000} negatively charged muons were propagated through various materials using \geant{} version 10.5. Production rates are shown separately for muon initial kinetic energies \SI{280}{GeV} \subref{fig:n_z:a} and \SI{1000}{GeV} \subref{fig:n_z:b}. The muon starting point was at \SI{-2000}{g/cm^2}. Note that the neutron yield in lead was scaled down by a factor of 2 in \subref{fig:n_z:a}.}
    \label{fig:n_z}
\end{figure}

Muons of specified kinetic energies were propagated through a box made from various reference materials with square front face with \SI{2000}{g/cm^2} lateral dimension 
and \SI{4000}{g/cm^2} in length. 
The muon propagation started at the centre of the front face and the initial momentum pointed along the long axis of the box. Produced neutrons were counted and special care was taken not to double-count neutrons after inelastic scattering, for which \geant{} terminates the track of the initial neutron and treats all the final-state ones as new particles.\footnote{In our counting of produced neutrons, we discarded the first daughter neutron as created by \geant{} in the inelastic scattering process.} In order to allow the neutron production to reach equilibrium, neutrons were counted only between \SI{1000}{g/cm^2} and \SI{3000}{g/cm^2} along the long axis. Figure~\ref{fig:n_z} shows how the yield stabilizes within the first few \SI{100}{g/cm^2} of various materials. Yield variations along the muon track are due to the small-number statistics of large cascades, each containing many neutrons. These variations contribute substantially to the uncertainty of neutron yield calculations.

\begin{figure}[!htb]
  \centering
    \begin{subfigure}{0.49\textwidth}
   \centering
   \includegraphics[width=0.9\stdfigwidth]{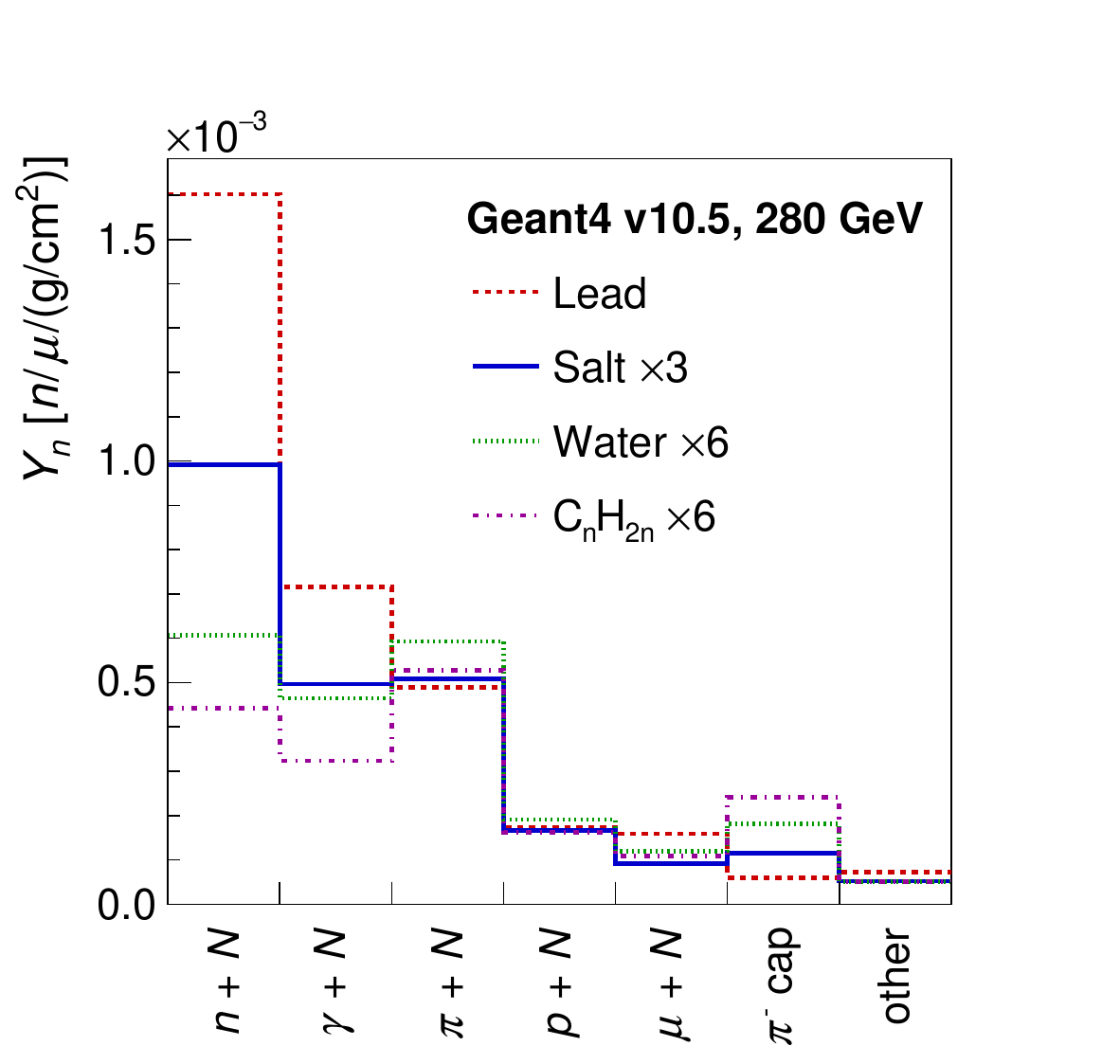}
    \caption{}
    \label{fig:g4_n_sources_1d:materials}
  \end{subfigure}
  \begin{subfigure}{0.49\textwidth}
  \centering
    \includegraphics[width=0.9\stdfigwidth]{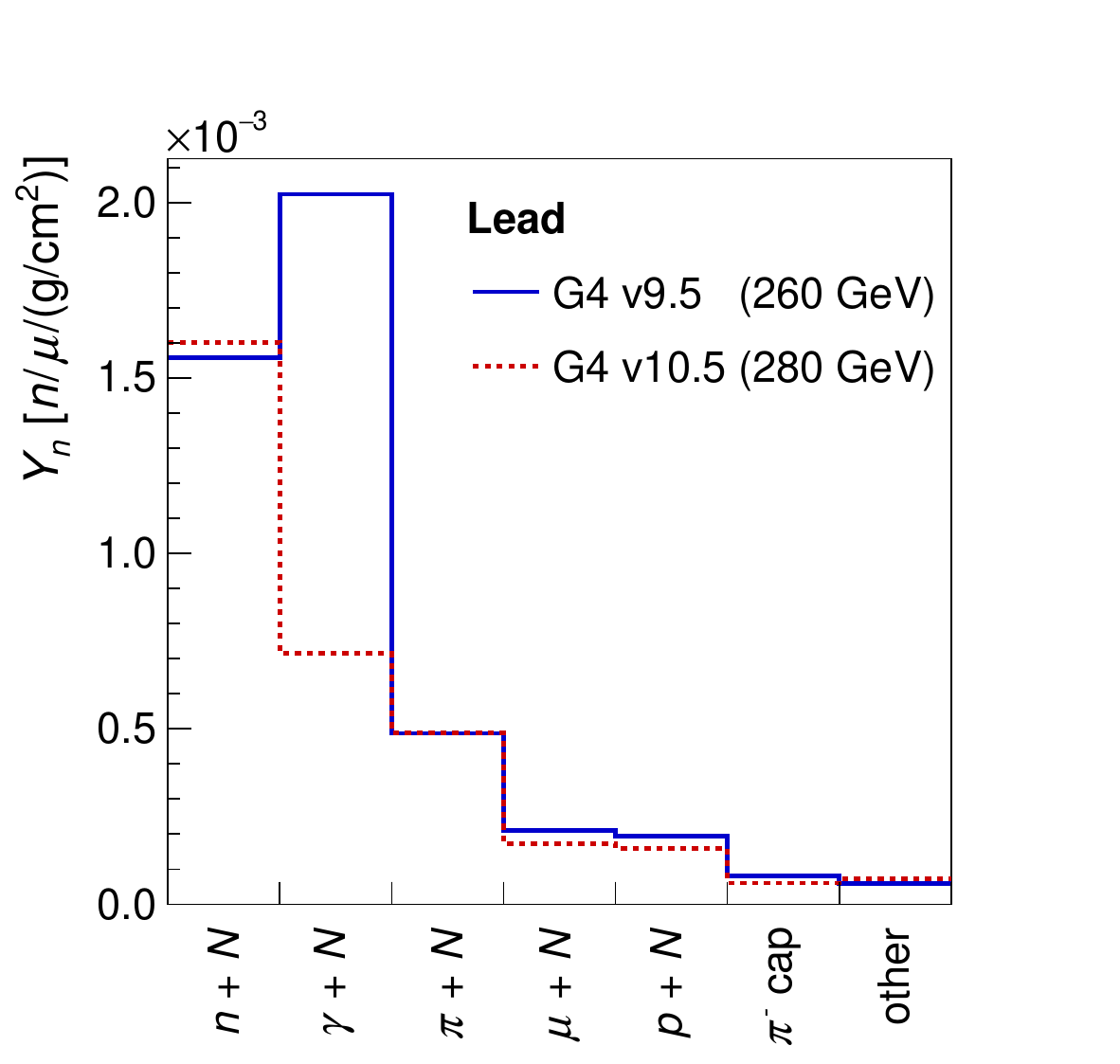}
    \caption{}
    \label{fig:g4_n_sources_1d:versions}
  \end{subfigure}
  %
  \caption[Neutron production sources.]{Neutron yield by production process.
  \protect\subref{fig:g4_n_sources_1d:materials}~Contribution of various processes to neutron production for 280\,GeV muons in different materials as simulated in \geant{} version 10.5. Note that the yields of lighter materials are scaled up in order to be visible. \protect\subref{fig:g4_n_sources_1d:versions}~Results of simulations of 260\,GeV and 280\,GeV muons in lead with \geant{} versions 9.5 and 10.5, respectively.
  The dominant processes are: neutron inelastic ($n + N$), photo-nuclear ($\gamma + N$), pion inelastic ($\pi + N$), muon nuclear ($\mu + N$), proton inelastic ($p + N$), and nuclear capture of $\pi^-$ at rest ($\pi^{-} \text{cap}$).}
  \label{fig:g4_n_sources_1d}
\end{figure}

\begin{figure}[!htb]
  \centering
  %
  \begin{subfigure}{0.49\textwidth}
   \centering
   \includegraphics[width=1.1\stdfigwidth]{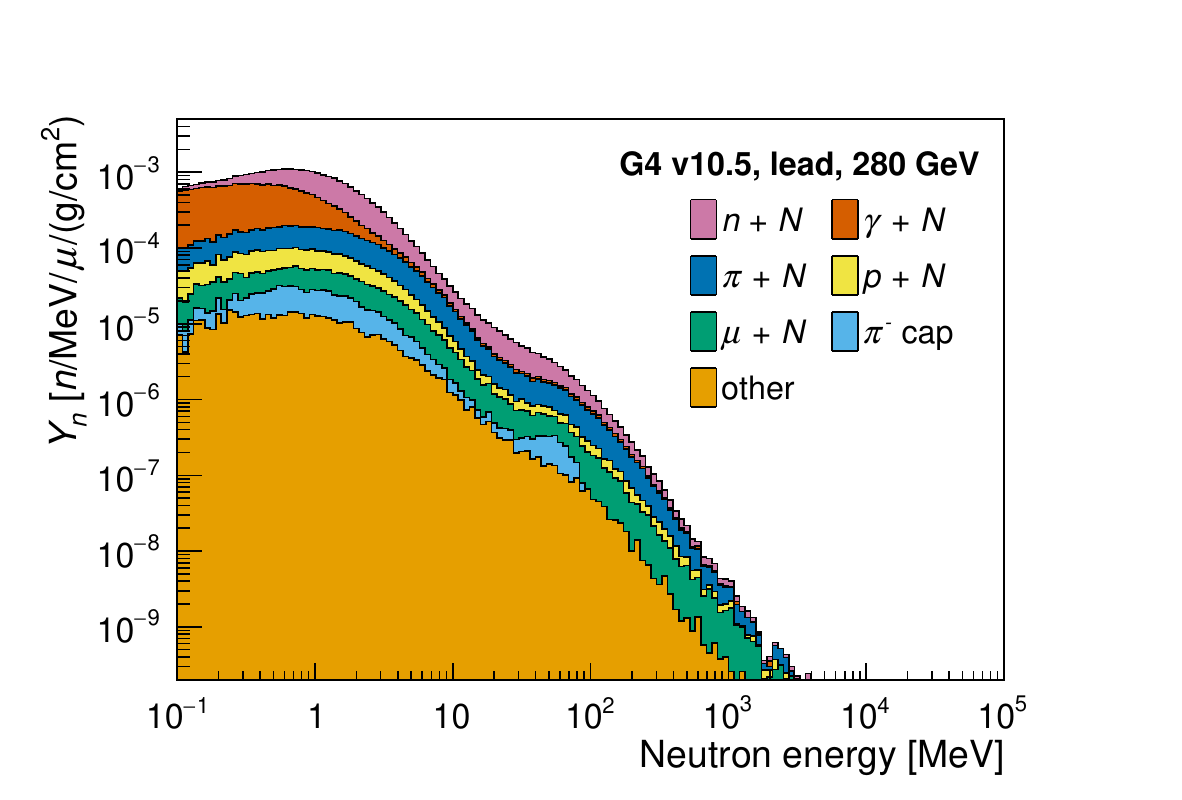}
    \caption{}
    \label{fig:g4_en_n_sources:g410}
  \end{subfigure}
  \begin{subfigure}{0.49\textwidth}
  \centering
    \includegraphics[width=1.1\stdfigwidth]{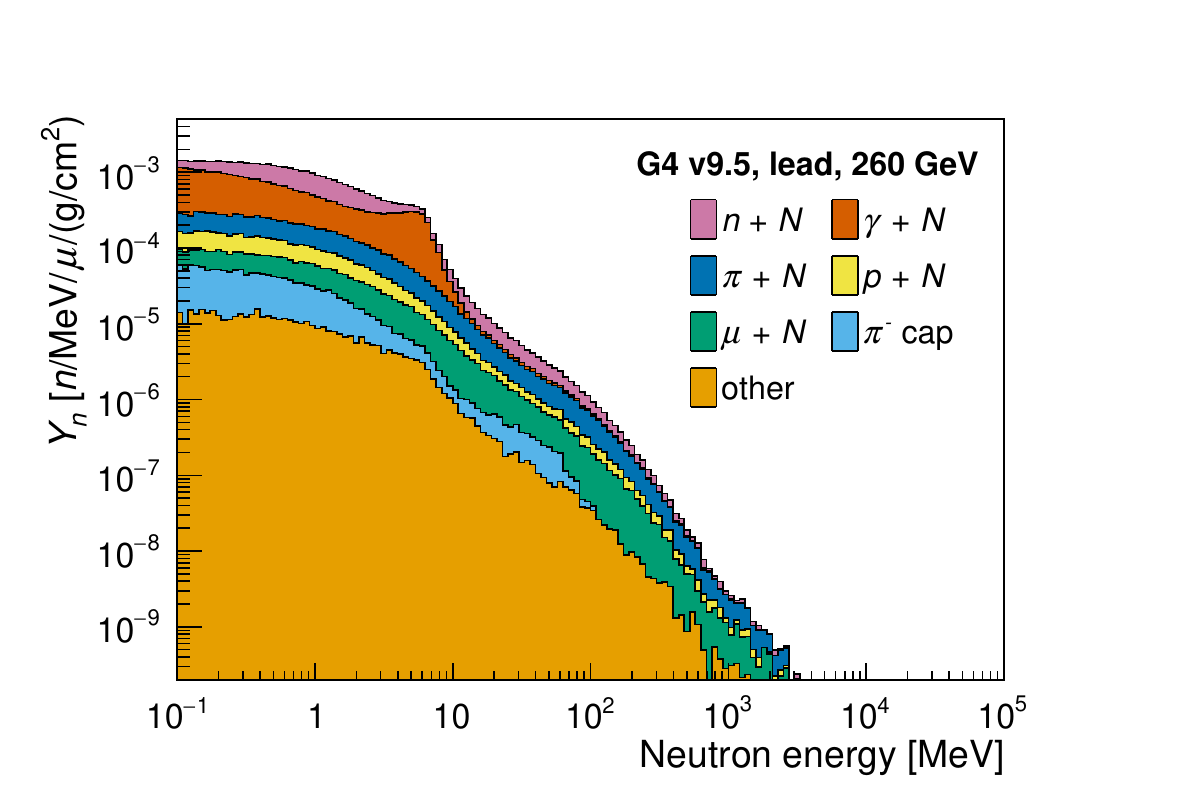}
    \caption{}
    \label{fig:g4_en_n_sources:g495}
  \end{subfigure}
  \caption{Energy spectra of neutrons produced in lead from different production processes. Production spectra \protect\subref{fig:g4_en_n_sources:g410} from 280\,GeV muons were simulated with \geant{} version 10.5 and \protect\subref{fig:g4_en_n_sources:g495} from 260\,GeV muons were simulated with version 9.5. The dominant processes are described in the caption of Fig.~\ref{fig:g4_n_sources_1d}.}
  \label{fig:g4_en_n_sources}
\end{figure}

Based on our simulations with \geant{} version 10.5, the largest contributor to the neutron production in heavier materials is the neutron inelastic scattering off a nucleus while for lighter materials, this process has contribution similar to the nuclear photo-production and pion inelastic scattering. Figure~\ref{fig:g4_n_sources_1d:materials} compares neutron yields for individual processes in different materials for \SI{280}{GeV} muons. Surprisingly, the larger number of electromagnetic cascades, expected to be produced in lead as compared to the lighter materials, does not result in an enhanced production of neutrons from gammas in version 10.5. We compared these results with simulations with \geant{} version 9.5, shown in Figure~\ref{fig:g4_n_sources_1d:versions}. There is marked reduction in the gamma-induced production in version 10.5 which cannot be explained by the small difference (20\,GeV) in initial muon energies. Energy spectra of neutrons from individual processes for lead are shown in Fig.~\ref{fig:g4_en_n_sources}. It can be seen that the significant reduction in neutron yield from photon interactions in version 10.5 is responsible for the change in the spectrum at a few MeV. We can conclude that the smaller neutron yield from gammas in version 10.5 is responsible for the reduction in the total neutron yield when compared to version 9.5. The exact reason for this change is not clear and further investigation goes beyond the scope of the current work.

\begin{figure}[!htb]
  \centering
  \begin{subfigure}{0.49\textwidth}
    \tikz{%
    \node () {\includegraphics[width=\stdfigwidth]{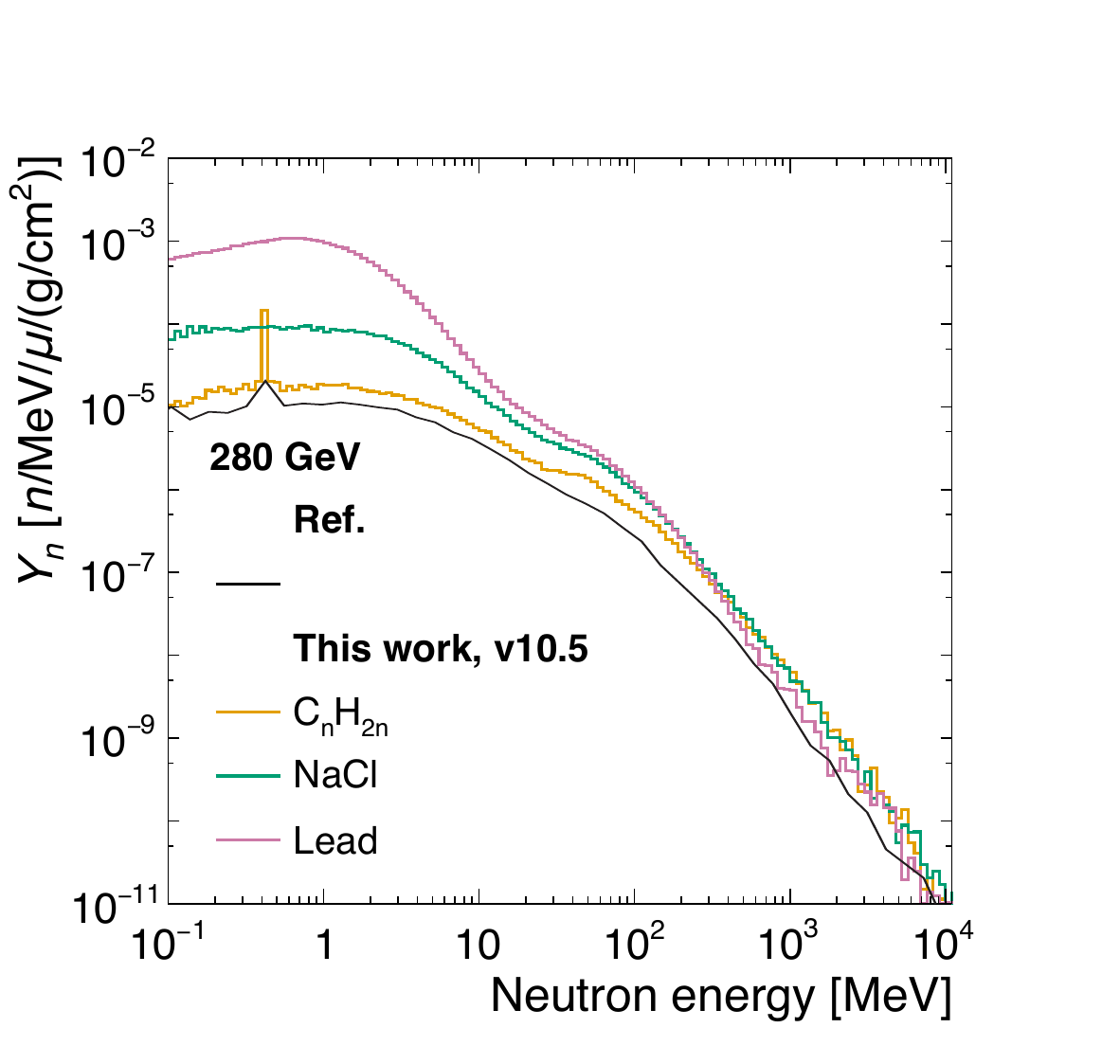}};
    \node[right] () at (-1.45cm,0.05cm) {\footnotesize\textsf{\textbf{\cite{al2009}, v8.2}}};
    \node[right] () at (-2.05cm,-0.38cm) {\footnotesize\textsf{C$_\text{n}$H$_\text{2n}$}};
    }
    \caption{}\label{fig:n_energy:ch2}
  \end{subfigure}
  \begin{subfigure}{0.49\textwidth}
    \includegraphics[width=\stdfigwidth]{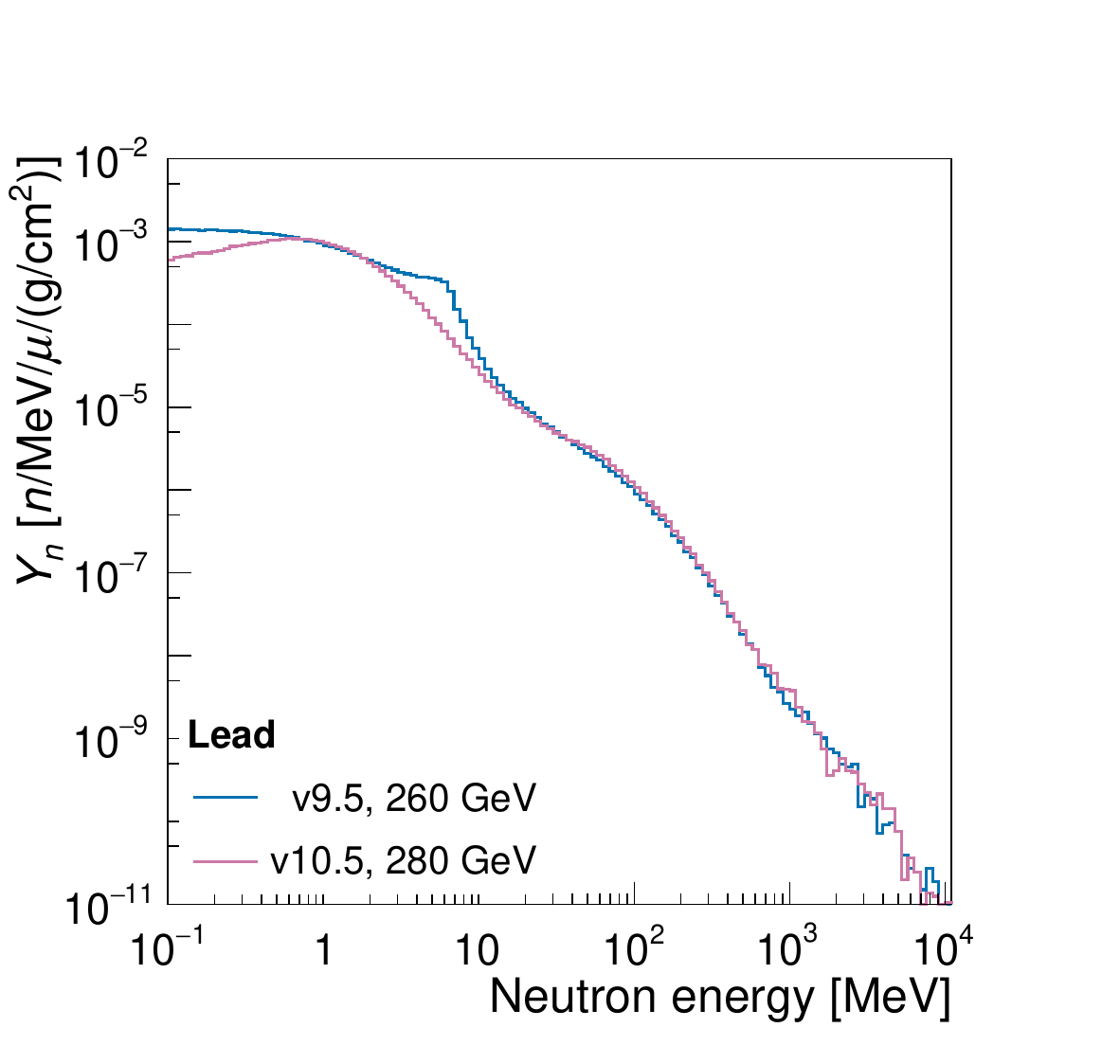}
  \caption{}\label{fig:n_energy:lead}
  \end{subfigure}
  \caption{Energy spectrum of muon-induced neutrons. \protect\subref{fig:n_energy:ch2} Simulations with  \geant~version 10.5 of 280\,GeV muons in \polyethylene, NaCl, and lead are compared, together with \geant~version 8.2 simulations in \polyethylene{} from Ref.~\cite{al2009}. The peak visible at 420\,keV for the \polyethylene{} sample is from the $\pi^-$ capture at rest on hydrogen, $\pi^- + p \rightarrow n + \pi^0$. \protect\subref{fig:n_energy:lead} Simulations of 260\,GeV muons in lead with \geant~version 9.5 are compared to simulations of 280\,GeV muons with version 10.5 of the toolkit.
  \label{fig:n_energy}}
\end{figure}

Final energy spectrum of neutrons in three different materials is shown in Fig.~\ref{fig:n_energy:ch2}. The neutron production in lead is enhanced compared to the lighter materials but this enhancement is substantial mainly at lower neutron energies (below about 50\,MeV), whereas the difference in spectral shapes and the absolute neutron yields becomes less significant for larger neutron energies, as had already been reported in Ref.~\cite{al2009}. The spectrum in polyethylene (\polyethylene) is compared with the results of simulations with \geant{} version 8.2 presented in Ref.~\cite{al2009} and the new version gives about 1.5 greater total yield. Figure~\ref{fig:n_energy:lead} compares spectra in lead as simulated in \geant{} versions 10.5 and 9.5. The changes in neutron production between the two versions of the toolkit were discussed in the previous paragraph.

\begin{figure}[!htb]
    \centering
    \begin{subfigure}{0.49\textwidth}
    \includegraphics[width=\stdfigwidth]{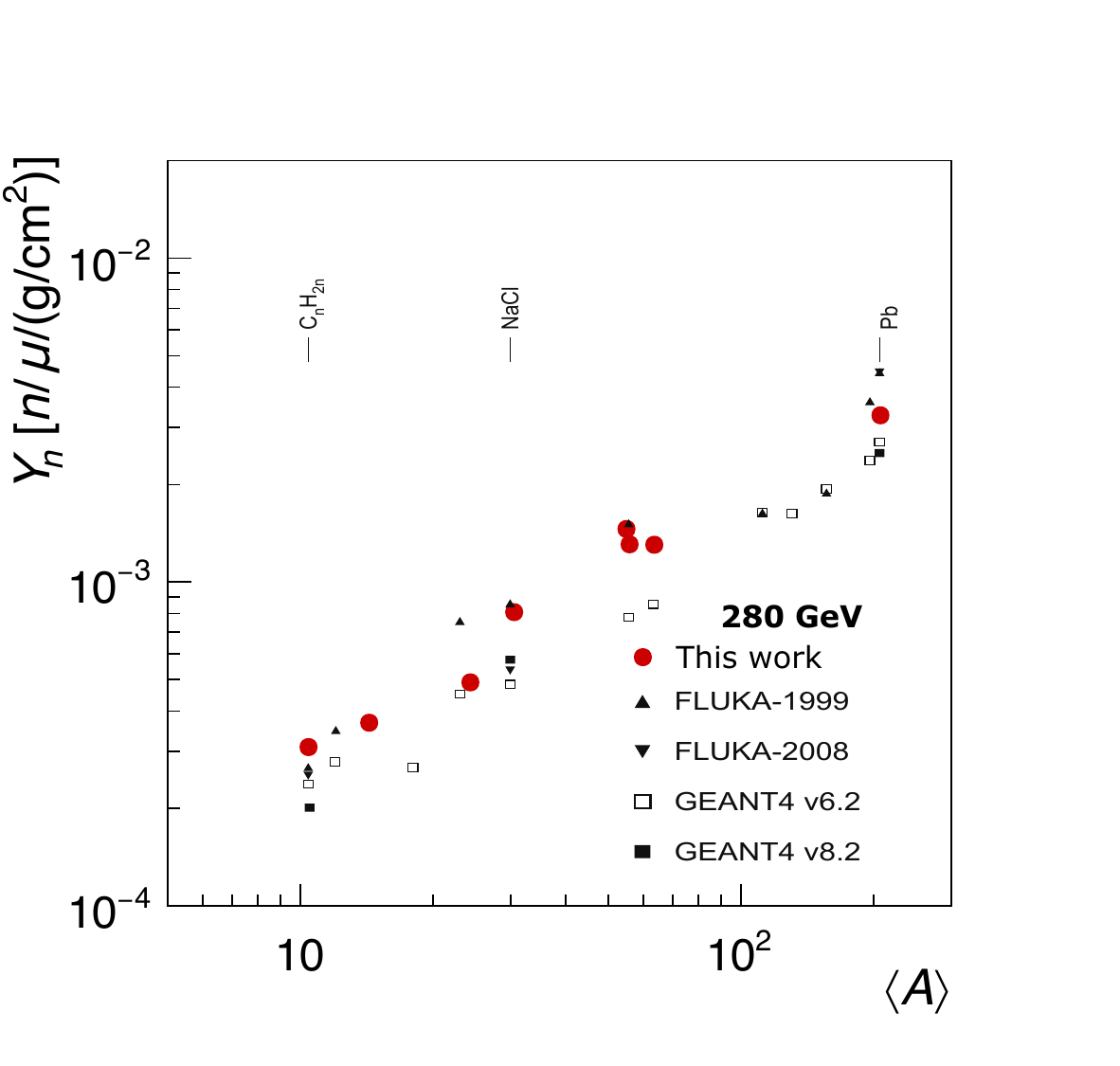}%
    \caption{}\label{fig:n_yield:a}
    \end{subfigure}%
    \begin{subfigure}{0.49\textwidth}
    \includegraphics[width=\stdfigwidth]{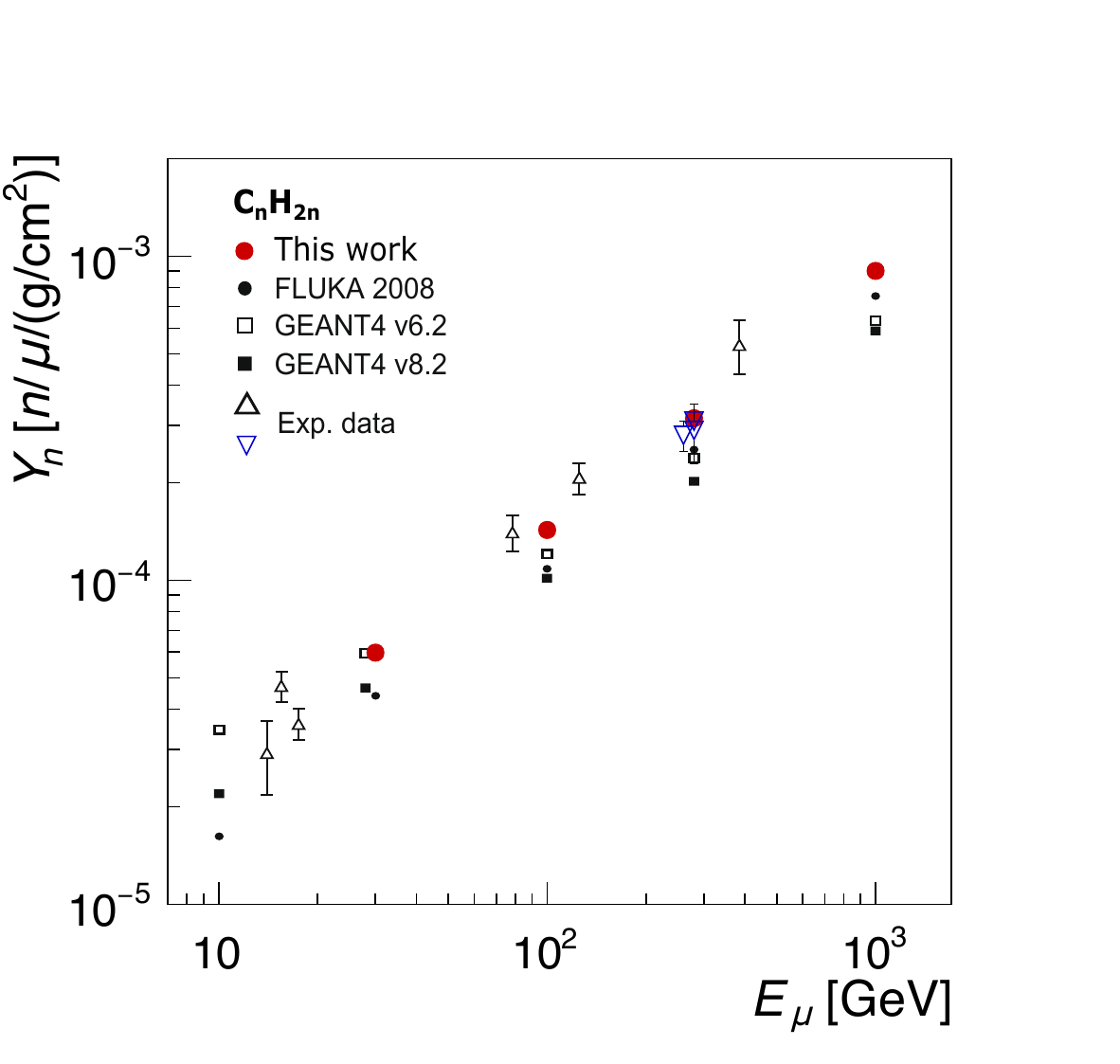}
    \caption{}\label{fig:n_yield:e}
    \end{subfigure}
    \caption{Neutron yields as functions of the mean atomic weight $\left<A\right>$ of the target material (for 280\,GeV muons) \protect\subref{fig:n_yield:a} and of the initial muon energy \protect\subref{fig:n_yield:e} (for polyethylene). In \protect\subref{fig:n_yield:a}, current simulations with \geant~version 10.5 are compared with \fluka{} and \geant{} simulations from Ref.~\cite{al2009}.
    In \protect\subref{fig:n_yield:e}, our simulations are compared to previous simulations from Refs.~\cite{al2009,ha2005}. Measurements of neutron yields from cosmic muons in organic scintillator at various depths are included as black up-triangles for older measurements in Refs.~\cite{Boehm2000,Hertenberger1995,Bezrukov1973,Enikeev1987,Aglietta1989} and as blue down-triangles for more recent and more accurate measurements at similar depths in Refs.~\cite{borexino2013,kamland2010,lvd2013}.}
    \label{fig:n_yield}
\end{figure}

We also studied the dependence of neutron yield on the type of material and on the muon energy for our nominal \geant{} version 10.5. Figure~\ref{fig:n_yield} shows the dependence of the total neutron yield on the atomic weight for several materials (Fig.~\ref{fig:n_yield:a}) and the yield dependence on the initial muon energy in polyethylene (C$_n$H$_{2n}$) (Fig.~\ref{fig:n_yield:e}). The results are compared to the simulations reported in Ref.~\cite{al2009} where various simulation software packages were used. Comparison between the \geant{} version 10.5 and previous versions shows continuous development of the neutron production models resulting in a noticeable change in neutron yield across multiple materials and muon energies. For the yield dependence on muon energy, Fig.~\ref{fig:n_yield:e}, we included data points from available measurements in a scintillator with a chemical formula similar to \polyethylene{} \cite{Boehm2000,Hertenberger1995,Bezrukov1973,Enikeev1987,Aglietta1989,borexino2013,kamland2010,lvd2013}. A good agreement is seen between data and our simulations for muon energy close to 280\,GeV, equivalent to the mean cosmic-ray muon energy at the depths relevant to this work.

We summarize neutron yields in polyethylene (or scintillator of similar composition) and lead as simulated with different versions of \geant{} in Tab.~\ref{tab:n_yields_ref}. We also add measurements of yields. It can be seen that the simulated yields in polyethylene agreed among each other within 35\% in the older software and that it has approached the measured yield which varied only within 10\%. We need to note here that Refs.~\cite{borexino2013,lvd2013,kamland2010} evaluated neutron yield for the whole spectrum of muon energies corresponding to the depths in question, while we studied neutron yield for fixed muon energy. Simulated yield in lead varied more substantially. 
Two experimental data sets, \cite{ha2008,lr2013}, for depths relevant to this work, give significantly different neutron yields even when interpreted with the same \geant{} version. Also, data from \cite{ha2008} (\SI{1.31E-03}{\nfluxunit}) were reinterpreted with newer version of \geant{} \cite{lr2013} with a different result (\SI{3.40E-03}{\nfluxunit}).

We note that the interpretation of data in terms of neutron yield is complicated and is dependent on full Monte Carlo simulations of the experimental setup, including muon propagation, development of cascades, neutron production and detection (via thermal neutron capture). In the measurements mentioned in the previous paragraph, it was assumed that the discrepancy between measurements and simulations was solely due to the modelling of muon induced neutron production. It is clear that continuous development and improvements in \geant{} affect many models, not just those which affect neutron production, and it will affect the interpretation of measurements of neutron production yields. The direct comparison between previous measurements and current simulations is not straightforward. However, re-analysis of previous measurements informed by the newer version of \geant{} is beyond the scope of this work.

In conclusion, there seems to be better agreement for lighter elements than for lead where the measured neutron yields quoted in Tab.~\ref{tab:n_yields_ref} differ by a factor of up to 2.5 from our nominal simulations. Our geometry contains mainly light elements and heavy targets like lead are unlikely to make up a significant fraction of future experiment's construction materials (with the exception of xenon itself, but the neutron production in xenon can be easily tagged). We can use a factor of 2 as a conservative estimate of the systematic uncertainty in cosmogenic neutron production in our simulations described in the next section.

\newcommand{\mymulti}[2][1.2cm]{\multirow{2}{*}{\parbox{#1}{#2}}}

\begin{table}[!tb]
    \caption{Neutron yields in simulations with different versions of \geant{} (and one instance of \fluka{} simulation) for \polyethylene{}(or scintillator of similar composition) and lead for similar muon energies. The simulated neutron yield is shown in the third column. The ratio of the referenced simulations to our work is in column four.
    Neutron yields from data interpretation, where available, are included in column five. Column six includes the ratio of data and simulations from the referenced work.}
    \label{tab:n_yields_ref}
  \small
  \centering
  \sisetup{round-mode=places, round-precision=1}
  \sisetup{round-mode=figures,round-precision=3}
  \begin{tabular}{@{}r@{}l>{\centering\arraybackslash}p{1.cm}>{\centering\arraybackslash}p{2.2cm}>{\centering\arraybackslash}p{2cm}>{\centering\arraybackslash}p{2.8cm}>{\centering\arraybackslash}p{1.9cm}@{}}
    \toprule
      \multicolumn{2}{c}{G4 version}     & $E_\mu$ & Yield \par   [\nfluxunit] & Ratio to our work & Interpreted data [\nfluxunit]
                                                                                                & Data/Sims.                   \\
    \midrule
      \multicolumn{7}{@{}l}{\bf Scintillator/\polyethylene }                                                                   \\
                    8.2 & \cite{al2009}               & 280                  & \num{2.00E-04} & 0.65  & -               & -    \\
          \fluka{} 2006.3b & \cite{kamland2010}          & 260                  & \num{2.34E-04} & 0.76  & \num{2.80E-04}  & 1.19 \\ 
                    9.3 & \cite{lvd2013}              & 280                  & \num{2.17E-04} & 0.70  & \num{2.90E-04}  & 1.34 \\
                    9.6 & \cite{borexino2013}         & 283                  & \num{3.01E-04} & 0.97  & \num{3.10E-04}  & 1.03 \\
    {\bf Our work} 10.5 &                             & 280                  & \num{3.10E-04} & -     & -               & -    \\
    \midrule
      \multicolumn{7}{@{}l}{\bf Lead }                                                                                         \\
                    8.2 & \cite{ha2008}               & \multirow{4}{*}{260} & \num{2.37E-03} & 0.72  & \num{1.31E-03}  & 0.55 \\
                    9.5 & \cite{lr2013}               &                      & \num{4.59E-03} & 1.40  & \num{5.78E-03}  & 1.26 \\
                    9.5 & \cite{ha2008}/\cite{lr2013} &                      & \num{4.59E-03} & 1.40  & \num{3.40E-03}  & 0.74 \\
    {\bf Our work} 10.5 &                             &                      & \num{3.27E-03} & -     & -               & -    \\
    \bottomrule
  \end{tabular}%
\end{table}

A comparison of neutron capture rate in muon events has also been reported in \cite{exo}. An agreement within 40\% has been found between data and simulations using \fluka{} and \geant{} for all isotopes, supporting our conservative approach for systematic uncertainty.

One aspect of neutron production worth mentioning is the delayed neutron emission after activation of materials by muon-induced showers. This process has previously been observed in scintillators \cite{kamland2010,borexino2013,dayabay} and is also included in the physics of \geant{}, but has not been commonly discussed in the context of dark matter experiments. In addition to the scintillator, there are other materials used in detector components which are susceptible to emit delayed neutrons, in particular polytetrafluoroethylene (PTFE) used as reflective material which contains fluorine. Our simulation with \geant{} predicts cosmogenic production of $^{17}$N from $^{19}$F. The $^{17}$N radioisotope undergoes $\beta$-decay with a half-life of \SI{4.2}{s} to the metastable state $^{17*}$O, which then promptly decays to $^{16}$O emitting the neutron.
We determined the neutron yield in PTFE from 280\,GeV muons to be \SI{0.65e-3}{\nfluxunit}, of which 0.66\% comes from the delayed emission mechanism. We have summarised our calculated neutron yields in various materials in Table~\ref{tab:n_yields}.

\begin{table}[!htb]
    \caption{Neutron yields in various materials for an initial muon energy of 280\,GeV as simulated with \geant{}~version 10.5. The stated errors represent statistical uncertainty estimated by dividing the simulated dataset into smaller samples and it is driven by variations in neutron production along the muon path due to large cascades (which is also reflected in Fig.~\ref{fig:n_z}).}
    \label{tab:n_yields}
    \centering
    \small
    \begin{tabular}{lr@{$\pm$}llr@{$\pm$}l}
        \toprule
        Material & \multicolumn{2}{c}{Neutron yield} &    Material & \multicolumn{2}{c}{Neutron yield} \\
                & \multicolumn{2}{c}{[$\times$\SI{e-3}{\nfluxunit}]} &
                & \multicolumn{2}{c}{[$\times$\SI{e-3}{\nfluxunit}]} \\
        \midrule
        C$_n$H$_{2n}$     & \phantom{00000}\num[]{0.31} & 0.01 & Mg  &                \num[]{0.49} & 0.02 \\
        H$_{2}$O          &                \num[]{0.37} & 0.01 & Ti  & \phantom{00000}\num[]{1.39} & 0.06 \\
        polyhalite        &                \num[]{0.46} & 0.02 & Mn  &                \num[]{1.46} & 0.04 \\
        PTFE              &                \num[]{0.65} & 0.03 & Fe  &                \num[]{1.31} & 0.05 \\
        NaCl              &                \num[]{0.81} & 0.03 & Cu  &                \num[]{1.30} & 0.05 \\
                          &                \multicolumn{2}{c}{}& Pb  &                \num[]{3.27} & 0.13 \\
        \bottomrule
    \end{tabular}
\end{table}

\section{Muon background in a next-generation liquid xenon experiment}\label{bkgsim}

We carried out simulations to determine the rate of potential background events caused by cosmic-ray muons in a next generation dark matter experiment operating at a depth of around 3\,k\mwee. The main detector is a dual-phase xenon time projection chamber (hereafter LXe-TPC) containing 70\,tonnes of active liquid xenon (LXe), corresponding to a $\sim$10-fold upscale of the existing experiments LZ \cite{LZexp2019} and XENONnT \cite{Aprile2020}.
Our main case study is the existing site at the Boulby Underground Laboratory (UK) at a depth of 2850\,\mwe, with a muon flux of $3.75\times10^{-8}$~cm$^{-2}$s$^{-1}$~\cite{lr2013} (this flux is very similar to that at the LNGS in Italy). A potential, deeper location was also investigated at a depth of 3575\,\mwe\ with an estimated muon flux of $1.13\times10^{-8}$~cm$^{-2}$s$^{-1}$.

\subsection{Simplified geometry model}\label{model}

\begin{figure}[!ht]
    \centering
    \begin{subfigure}{0.32\textwidth}
    \includegraphics[height=5.5cm,trim=12cm 4.6cm 13.5cm 4.8cm,clip]{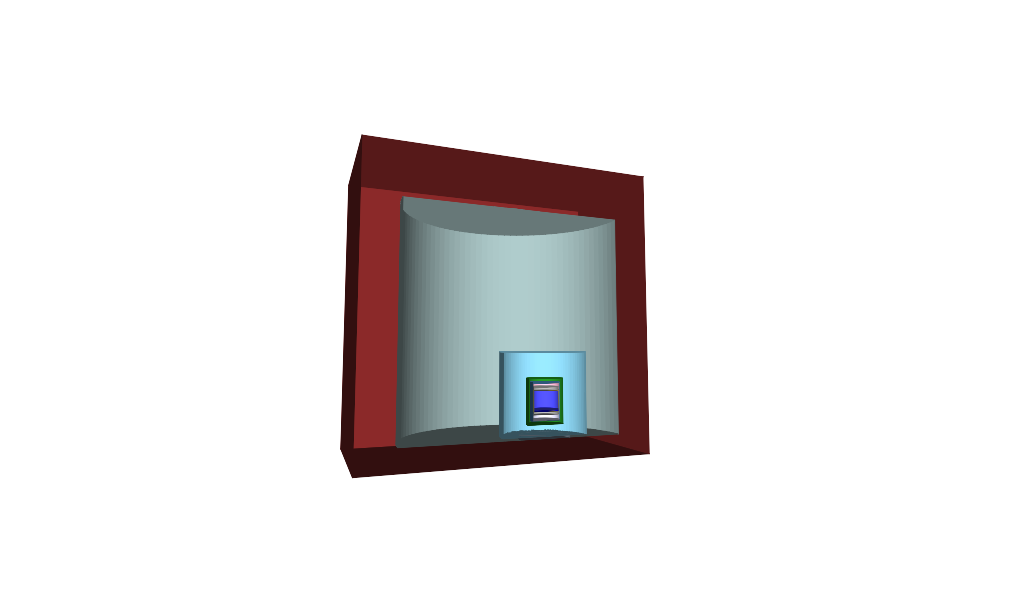}%
    \caption{\label{fig:geom:full}}
    \end{subfigure}
    \begin{subfigure}{0.6\textwidth}
    \includegraphics[height=5.5cm,trim=3.8cm 2.5cm 5cm 2.5cm,clip]{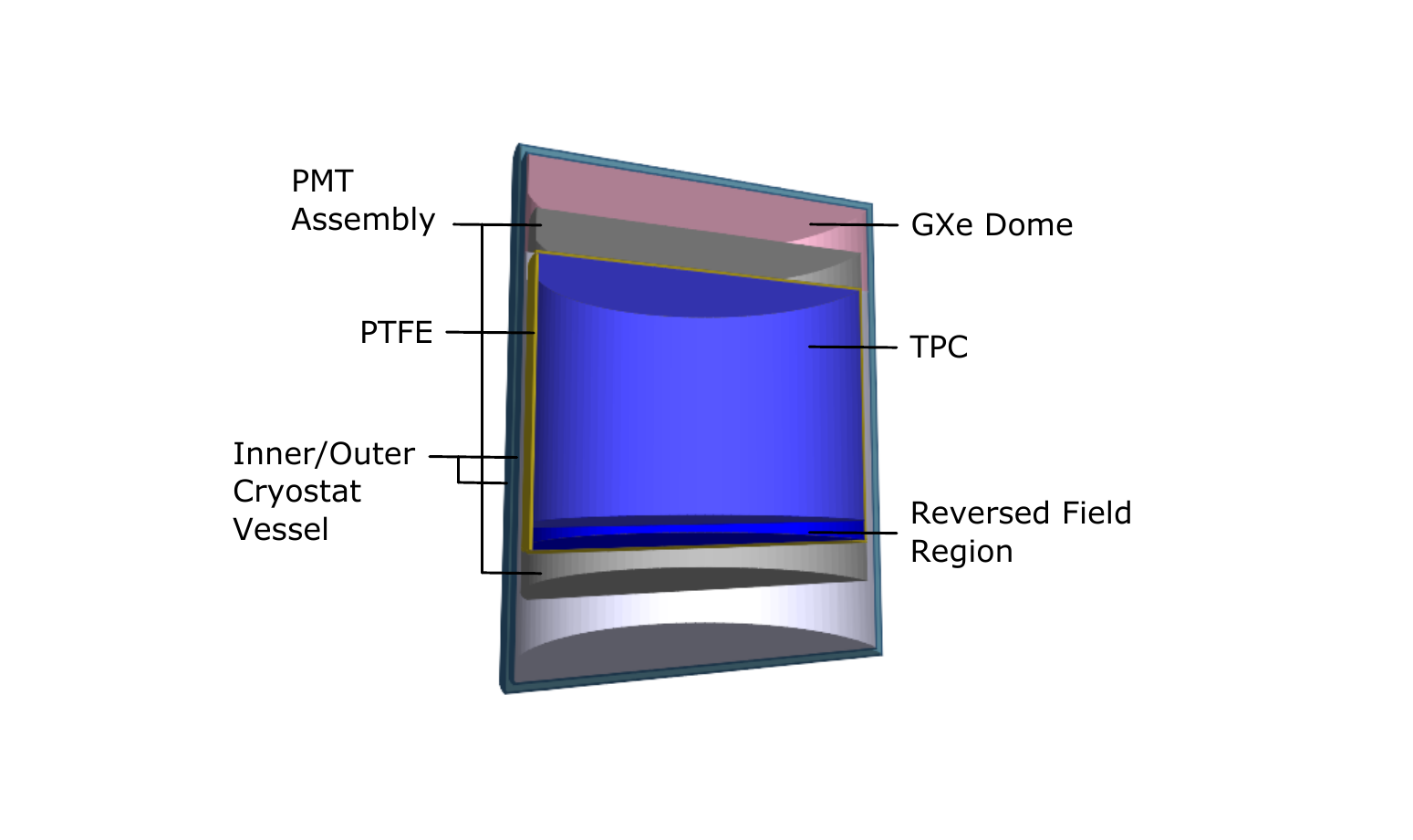}
    \caption{\label{fig:geom:cryo}}
    \end{subfigure}
    \caption{Visualisations of the simplified geometry model used in the simulations. \protect\subref{fig:geom:full} Cross-sectional view of full cylindrical cavern (gray) surrounded by rock material (dark red). The water tank (cyan) containing the detector is off-set from the cavern centre. The detector is enveloped with a layer of scintillator (green). \protect\subref{fig:geom:cryo} Labeled cross-sectional view of the detector cryostat.}
    \label{fig:geom}
\end{figure}

A simplified experimental hall and detector geometry model was used in simulations and is shown in Fig.~\ref{fig:geom}. 
The main elements of the experiment were a vacuum cryostat approximately 4\,m in diameter and 5\,m in height containing the xenon detector, an anti-coincidence veto system surrounding the main detector, all located within a water tank with 12\,m in diameter for shielding of local radioactivity backgrounds. The model was based loosely on the LZ and XENONnT designs \cite{LZexp2019,Aprile2020} and scaled to larger mass to meet the required sensitivity: a tenfold improvement over that of the current generation of liquid xenon experiments. The main ingredients to the design of the simulation were as follows:
\begin{itemize}
    \item The rock material around the cavern was included as this allowed starting the propagation of cosmic muons from within the rock. This ensured that production of high-energy cascades and fast neutrons in the rock that can propagate down to the shielding, veto system and the detector itself, was taken into account.
    \item All materials with significant mass which were expected to play a role in the particle production and propagation in and around the active part of the detector were included.
    \item The expected structure of the detection elements and the shielding was modelled to a certain level of detail (e.g.~a~water tank used to attenuate any external neutrons and gammas, a layer of liquid scintillator which is envisioned as an optional additional external veto system).
    \item A realistic layout of the space in the experimental cavern was considered: significant space is required above the water tank and the water tank is expected to be offset from the centre of the cavern in order to make efficient use of space for ancillary subsystems during installation and operation.
\end{itemize}

A visualisation of the full geometry model can be seen in Fig.~\ref{fig:geom:full}. The model included a cylindrical cavern (diameter and height of 30\,m) surrounded by rock. The detector was placed at the bottom of the cavern and offset from the centre by $\sim$4\,m. It consisted of a cylindrical cryostat containing the TPC and was surrounded by 50\,cm of liquid scintillator and placed within a water tank (WT) (12\,m diameter $\times$ 11\,m height). The water and the scintillator served as both shielding against external radiation and as active veto outside the TPC.

The total shielding thickness was informed by previous experience with the LUX and LZ experiments. The neutron flux from radioactivity in the rock can be efficiently attenuated by 1\,m of water for a multi-tonne (${>}$10~tonne) dark matter experiment. LZ simulations \cite{akerib2020,akerib2021} showed that 3\,m of water+scintillator shield are sufficient to attenuate $\gamma$-rays from rock to a level where this background can be neglected compared to other sources. To account for the higher sensitivity of the future experiment and, in particular, to decrease the background for a $0\nu\beta\beta$ search with $^{136}$Xe, we increased this thickness to 4\,m on all sides except below the detector. The water+scintillator thickness there was reduced to 2\,m and an additional 30\,cm layer of steel was placed beneath the water tank, providing the same total areal density of shielding. This reduction in height below the heavy cryostat and the scintillator containers will ease the design of the support structures.

The detector cryostat was approximated as a cylinder with an overall diameter of 3.9\,m and a height of 4.9\,m. The cryostat was made of two titanium vessels 2\,cm thick with 5\,cm of evacuated space in between. The inner cryostat vessel was filled with LXe up to the top of the TPC, topped by gaseous xenon (GXe). The active volume of the LXe-TPC (3.5\,m diameter $\times$ 2.5\,m height, between a cathode at the bottom and a gate grid just below the liquid surface; neither of which were included in the model) was enclosed by a 3\,cm thick PTFE `field cage'. (Note that the thickness of PTFE in the field cage will eventually be determined by the structural analysis of the TPC and the outgassing rate, and the current value translating to about 2.8\,tonnes of PTFE is unlikely to be adopted in a realistic design.) The active volume would be readout by two arrays of photomultiplier tubes (PMTs) located at the bottom and top of the field cage, in the liquid and gaseous phases, respectively. The arrays were modelled as two uniform volumes of steel with reduced density of \SI{0.4}{g/cm^3}, or about 5$\%$ of the standard density of steel, simulating metal components of the structure of the arrays and matching its mass. Other materials which often appear in such structures were neglected. In addition to the drift volume there was a separate volume of Reversed Field Region (RFR) at the bottom of the TPC. A thin layer (8\,cm) of LXe (`LXe skin') was kept between the TPC and the cryostat walls. This layer would be used as an additional anti-coincidence system based on detection of scintillation light, similar to the LZ design.
A closeup of the cross section of the modelled cryostat can be seen in Fig.~\ref{fig:geom:cryo}.
Dimensions of the main features of the geometry used in the simulations are summarized in Table~\ref{tab:parameters}. We note that this is not the proposed design of the next-generation experiment, but a simplified setup for the presented study.

\begin{table}[!ht]
    \centering
    \caption{Summary of the elements in the simulated geometry model. The elements are ordered hierarchically from the inner most to the all-including rock volume. Most of the elements were modeled as cylindrical and their diameter (D) and height (H) are listed. Where appropriate, thickness of a layer of the material is listed. For the rock volume, width of its bottom sides and the height are included in the D and H columns. The total amount of LXe contained in the geometry model was 108\,tonnes ($\rho_\text{Xe} = \SI{2.953}{g/cm^3}$), with 70\,tonnes in the TPC, 5.7\,tonnes in the RFR, 31\,tonnes in the skin (including 22.5\,tonnes in the bottom part of the cryostat). We note that this is not the proposed design of the next-generation experiment, but a simplified setup for the presented study.
    \label{tab:parameters}}
    \small
  \begin{tabular}{@{}l@{}lrrrl@{}}
    \toprule
    \multicolumn{2}{@{}l}{Volume} & D {[}m{]}     & H {[}m{]}        & Thickness         & Comments                            \\
    \midrule
    \multicolumn{6}{@{}l}{\textbf{Cylindrical}}                                                                                \\
    \phantom{m} & LXe-TPC      &  3.5\phantom{0}  &  2.5\phantom{0}  &                   &                                     \\
                & field cage (PTFE) &     \phantom{0}  &     \phantom{0}  & 3 cm              &                                     \\
                & RFR          &  3.5\phantom{0}  &  0.2\phantom{0}  &                   &                                     \\
                & PMT array    &  3.5\phantom{0}  &  0.4\phantom{0}  &                   & on top and at bottom of TPC         \\
                & LXe skin     &                  &                  & 8 cm/70 cm        & side/bottom                         \\
                & GXe          &  3.72            &  0.93            &                   & starts level with top PTFE          \\
                & Cryo Inner   &  3.72            &  4.76            & 2 cm              & inner D and H                       \\
                & Cryo Vacuum  &                  &                  & 5 cm              &                                     \\
                & Cryo Outer   &  3.9\phantom{0}  &  4.94            & 2 cm              & outer D and H                       \\
                & GdLS         &  4.9\phantom{0}  &  5.94            & 50 cm             &                                     \\
                & Water        & 11.9\phantom{0}  & 10.94            & 3.5 m/3.5 m/1.5 m & side/top/bottom                     \\
                & Steel Shield &  6.9\phantom{0}  &  0.3\phantom{0}  & 30 cm             & centred below WT                    \\
                & Hall         & 30\phantom{.00}  & 30\phantom{.00}  &                   & TPC off-set by 4.05 m from centre   \\
    \multicolumn{6}{@{}l}{\textbf{Box}}                                                                                        \\
                & Rock         & 40\phantom{.00}  & 40\phantom{.00}  & 5 m/7 m/3 m       & min.~amount of rock around the cavern,    \\
    \multicolumn{5}{c}{}                                                                 & side/top/bottom                     \\
    \bottomrule
  \end{tabular}
\end{table}

To study the dependence of the results on the rock composition and the size of the cavern, several sets of simulations were carried out. The rock around the lab in the nominal simulations was made of either salt (NaCl, for the existing Boulby lab site) or polyhalite (K$_2$Ca$_2$Mg(SO$_4$)$_4${}$\cdot$2H$_2$O), as appropriate for a deeper site at 1400\,m (3575~\mwe). In addition to the nominal cavern model specified above, an alternative geometry was simulated which included a smaller, cubic cavern with a side of 19\,m. Two samples of limited statistics with rock made of NaCl and CaCO$_3$ were simulated for this alternative geometry. No noticeable differences were found in the nuclear recoil (NR) spectra in the main LXe target between the different simulations. The background estimations reported here are results of analysis of the simulations with the nominal geometry and with salt and polyhalite as the rock materials.

\subsection{Simulation of cosmic-ray muons underground}\label{muons}

Distributions of primary energies and directions of cosmic-ray muons were calculated using the MUSIC and MUSUN codes \cite{music1997,music} (Ref.~\cite{music} describes the procedure and muon transport through rock down to the experimental site). Muons were sampled on the top and side surfaces of a 40\,m cube that surrounded the cavern such that they needed to travel through at least 7\,m of rock at the top of the cavern and through at least 5\,m of rock on the sides. 
Production of high-energy cascades and fast neutrons in the rock that could propagate into the cavern was expected to reach equilibrium with their absorption within that distance.

The rate of simulated muons was \SI{0.8759}{s^{-1}} for the existing Boulby site within salt at 2850\,\mwe\ vertical overburden. The mean muon energy and zenith angle were calculated as 261\,GeV and 30.6\degree{}, respectively. The surface profile was assumed to be flat in these simulations (in reality, variations in elevation up to 30\,m exist on the surface over areas of a few km$^2$) but the normalisation of the muon flux was done based on the measurements and the overall uncertainty is dominated by that from neutron production (see Section \ref{validation}). For the proposed deeper site in polyhalite at 3575\,\mwe\ vertical overburden, the same muon distributions were used, but the equivalent sampling rate was recalculated to be \SI{0.2625}{s^{-1}}.

Muon transport through the modelled experimental site was done using the \geant{} version 10.5 simulation toolkit. Physical processes were modelled according to the toolkit's modular physics list Shielding.
We have compared this version with other simulations and measurements as described in Section~\ref{validation}.

In total, 800~million muons were simulated for each rock material, salt and polyhalite. 
These numbers correspond to approximately 29\,years and 97\,years of live time of the experiment, respectively, accounting for the larger depth of the site in polyhalite.

\subsection{Analysis of simulated data}\label{analysis}

The expected WIMP signature in a typical dark matter experiment, and in a xenon-based experiment in particular, consists of a single scatter event at low energy, usually $\lesssim$50\,keV, in anti-coincidence with other detectors (veto systems), which is classified as a nuclear recoil using specific discrimination techniques. Here we assumed a nuclear recoil energy threshold of 1\,keV. For a proper analysis and interpretation of the results (limit setting, at the moment), usually the profile likelihood ratio technique is used for signal (and background) estimation, utilising probability density functions constructed from detailed signal models plus signal-free ancillary data. In this work we adopted instead a simple background counting technique with the potential (irreducible) background satisfying the signal conditions described above.

We analysed the simulation output in the following way. The detector response (i.e.~the digitised PMT waveforms resulting from the prompt and delayed scintillation signals from each energy deposition in the active volume) was considered only in terms of the characteristic times over which signals were collected and the equivalent energy thresholds in the respective active volumes -- LXe-TPC, LXe skin, liquid scintillator, water tank. Energy depositions by ionising particles in the LXe-TPC were summed over 1\,ms to accumulate interactions within the TPC over the realistic readout time similar to the maximum electron drift time. This is equivalent to the collection of all prompt and delayed signals within a single readout. (Note that potential background events stored in this way have later been generated again and analysed with a much better time resolution to remove multiple scatters.) We distinguished the depositions by their origin as xenon nuclear recoils, muon ionisation, electromagnetic activity, and others. The simulations and analysis procedure allowed reprocessing of selected events to follow closely individual interactions within the time window of 1\,ms. Energy depositions in the skin, liquid scintillator and water tank were summed over \SI{1}{\micro\second}, irrespective of their origin. This time window is close to the realistic time window from existing experiments for anti-coincidences between prompt signals from different systems to remove background events. 
We chose thresholds of 100\,keV, 200\,keV and 200\,MeV in the skin, liquid scintillator and water tank, respectively, to trigger veto signals. 
Summed depositions in the LXe-TPC were then tested for anti-coincidence with the veto signals by requiring no veto signal to be present within 0.5\,ms before or after any TPC signal. This time window was chosen to tag the delayed signals from neutron capture.
The depositions in the LXe-TPC by nuclear recoils were required to be larger than 1\,keV while all the other depositions were required to be below 10\,keV. These requirements gave us pre-selected candidates for the background events.
The threshold energies and the anti-coincidence window are summarised in the upper part of Table~\ref{tab:cuts}.

\begin{table}[!ht]
    \caption{Summary of criteria used to select background events. The top part of the table lists criteria used to filter down events based on energy and timing. The bottom part lists conditions applied to the events at the single-recoil level. Energy thresholds are listed for depositions in individual parts of the detector system, the TPC, the LXe skin, liquid scintillator (LS), and water tank (WT). Depositions in the TPC were treated separately for Xe nuclear recoils (NR), and all other sources (non-NR). An anti-coincidence time window was applied between depositions in the TPC and the other 3 volumes.}
    \label{tab:cuts}
    \centering
    \small
    \newcommand{\tmpcc}[1]{\multicolumn{1}{c}{#1}}
    \begin{tabular}{@{}lr@{\,}lcccc@{}}
      \toprule
      \multirow{2}{5em}{Threshold} & \multicolumn{2}{c}{TPC (NR)} & TPC (non-NR) & Skin     & LS       & WT          \\
      \cmidrule(lr){2-7}
                                   & 1             & keV          & 10\,keV      & 100\,keV & 200\,keV & 200\,MeV    \\
      Anti-coincidence             & $\pm$0.5      & \si{\ms}     &              &          &          &             \\
      \midrule
      Fiducial volume              & 5       & \multicolumn{5}{@{}l}{cm from LXe boundary}                                  \\
      Single recoil                & $>$1    & \multicolumn{5}{@{}l}{\si{\keV} (no other depositions $>$ 10\,keV)}   \\
      No other recoil              & $>$0.5  & \si{\keV}    &         &          &          &                        \\
      \bottomrule
    \end{tabular}
\end{table}

To be considered as a background to a WIMP search, events in the LXe-TPC were restricted to have only single nuclear recoils above 1\,keV, no other energy deposition above 10\,keV (this is a conservative cut since these depositions would be easily detected and the event rejected), and no other nuclear recoils of energy above 0.5\,keV (this would be identified as a multi-scatter nuclear recoil event and rejected). Since nuclear recoils from neutron scattering (neutrons originated from outside the TPC) tend to occur near the periphery, fiducialization helps significantly to remove a large fraction of the background events, and we required the nuclear recoils to happen further than 5\,cm from the boundary of the active volume (yielding fiducial mass of \SI{64}{t}). These applied cuts are summarised in the lower part of Table~\ref{tab:cuts}. In summary, events that passed the initial cuts (discussed in the previous paragraph and summarised in the upper part of Table~\ref{tab:cuts}), were re-processed and examined closely and they were considered as background events if they passed the cuts from the lower part of Table~\ref{tab:cuts}.

A geometry model without the presence of the liquid scintillator veto system was also investigated in order to determine whether this additional detector was needed to suppress backgrounds from cosmogenic neutrons. The same simulated data were used and the absence of the scintillator was emulated by treating both volumes, scintillator and water, as a single volume of `water' with the corresponding energy threshold.

\subsection{Results}\label{results}

Simulations of the cosmic-ray muons show that, in the case of the shallower location in NaCl, about 380~muons per day pass through the active TPC region while there's about 4900~muons per day passing through the water tank. For the deeper location in polyhalite the numbers are 115~muons per day and 1500~muons per day, respectively.

\begin{figure}[!htb]
    \centering
    \begin{subfigure}{0.49\textwidth}
        \includegraphics[height=6cm]{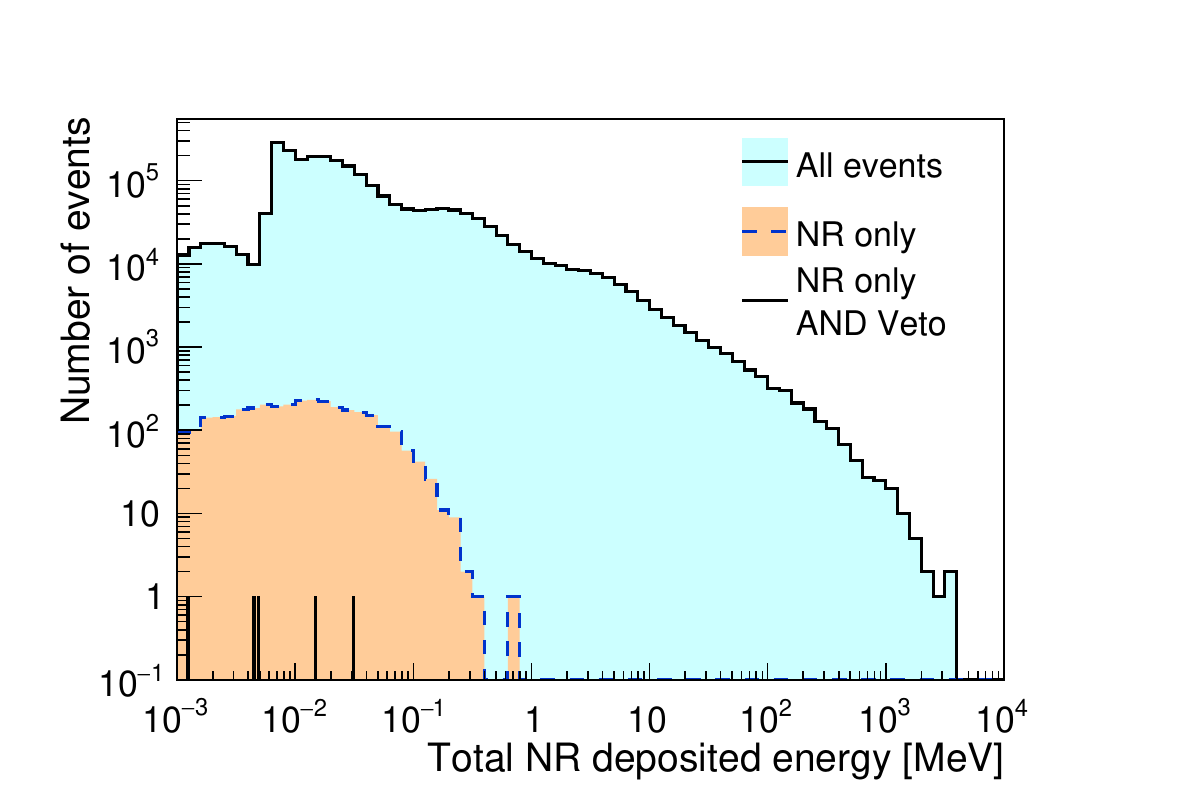}\hspace{-1cm}
        \caption{}\label{fig:spectra:tpc}
    \end{subfigure}%
    \begin{subfigure}{0.49\textwidth}
        \includegraphics[height=6cm]{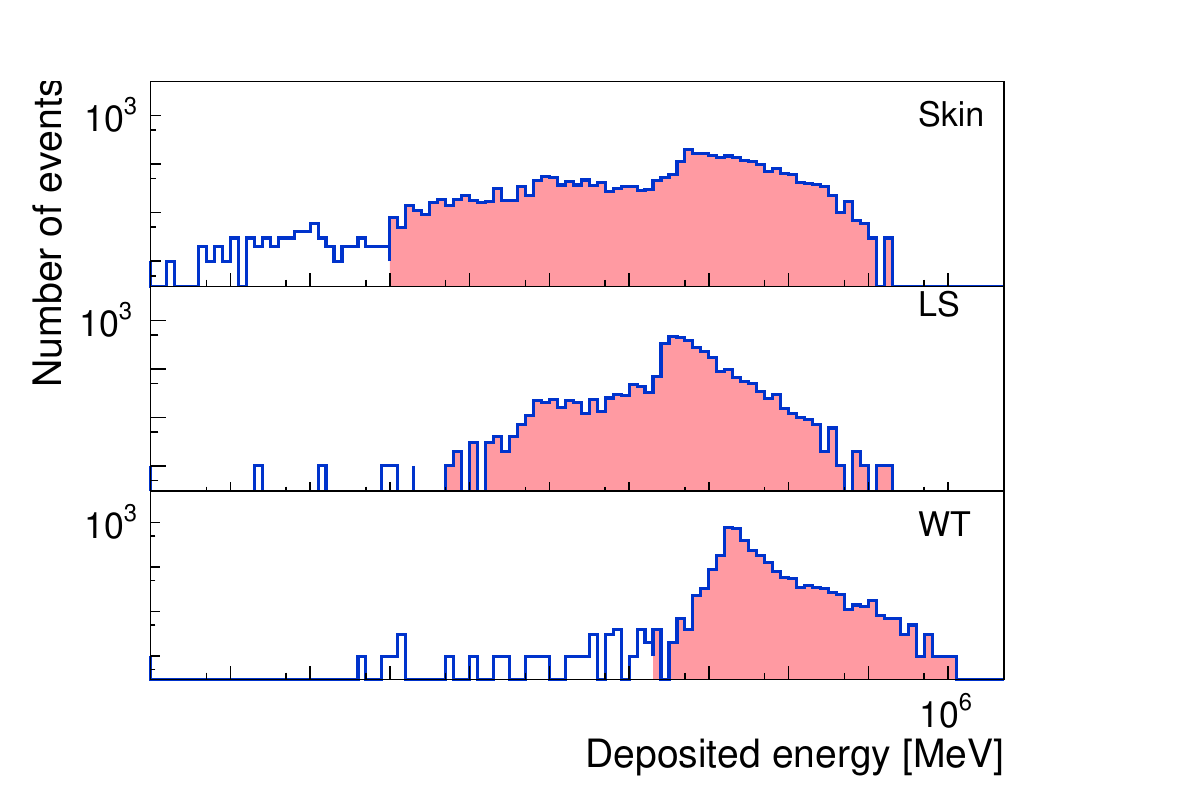}
        \caption{}\label{fig:spectra:veto}
    \end{subfigure} 
    \caption{\protect\subref{fig:spectra:tpc} Energy spectrum of energy depositions from nuclear recoils (NR) inside the TPC from the simulation with NaCl as the rock material. The solid histogram represents all simulated events in the sample with depositions from NR. The dashed histogram shows all events where energy depositions other than from NR were below the imposed threshold. Short vertical lines mark energies of 5 single events which also passed veto in the skin, scintillator, and water tank. Note the histograms are binned in logarithmic scale with 10~bins per decade in energy. \protect\subref{fig:spectra:veto} Energy spectra in the veto detectors: skin, liquid scintillator, and water tank. Events in coincidence with NR-only depositions in the TPC are included. Depositions above threshold are highlighted with the coloured area.}
    \label{fig:spectra}
\end{figure}

These muons generated neutrons that may cause unwanted backgrounds, as discussed above. Spectra of total energy depositions from nuclear recoils in the active volume of the TPC are shown in Fig.~\ref{fig:spectra:tpc} (for the standard 2850\,\mwe\ overburden with NaCl rock). The figure shows depositions in events without any selection requirements (most events also contained other energy depositions which were not included in the plotted energies) and in events where there were no depositions in the TPC other than from the recoils. The vertical black lines indicate 5 events with only nuclear recoils and without any coincident signals in the skin, liquid scintillator, or water volumes, i.e.~these are events passing the first part of the selection as described in Section~\ref{analysis} (also Table~\ref{tab:cuts}, upper part; no multiple scatter or fiducial volume cuts applied yet). The sharp rise in the number of events at about 10 keV for the histogram labeled `All events' is due to the nuclear recoils from muon Coulomb scattering where \geant{} `produces' a recoiling nucleus only above a certain energy threshold. This feature is not visible in the other spectra since events where muons had deposited energy inside the TPC were rejected. Figure~\ref{fig:spectra:veto} shows the spectra of events in the LXe skin (top), liquid scintillator (middle) and water (bottom) in coincidence with events in the TPC which have nuclear recoils only. 

\begin{figure}[!htb]
    \centering
    \includegraphics[height=6cm]{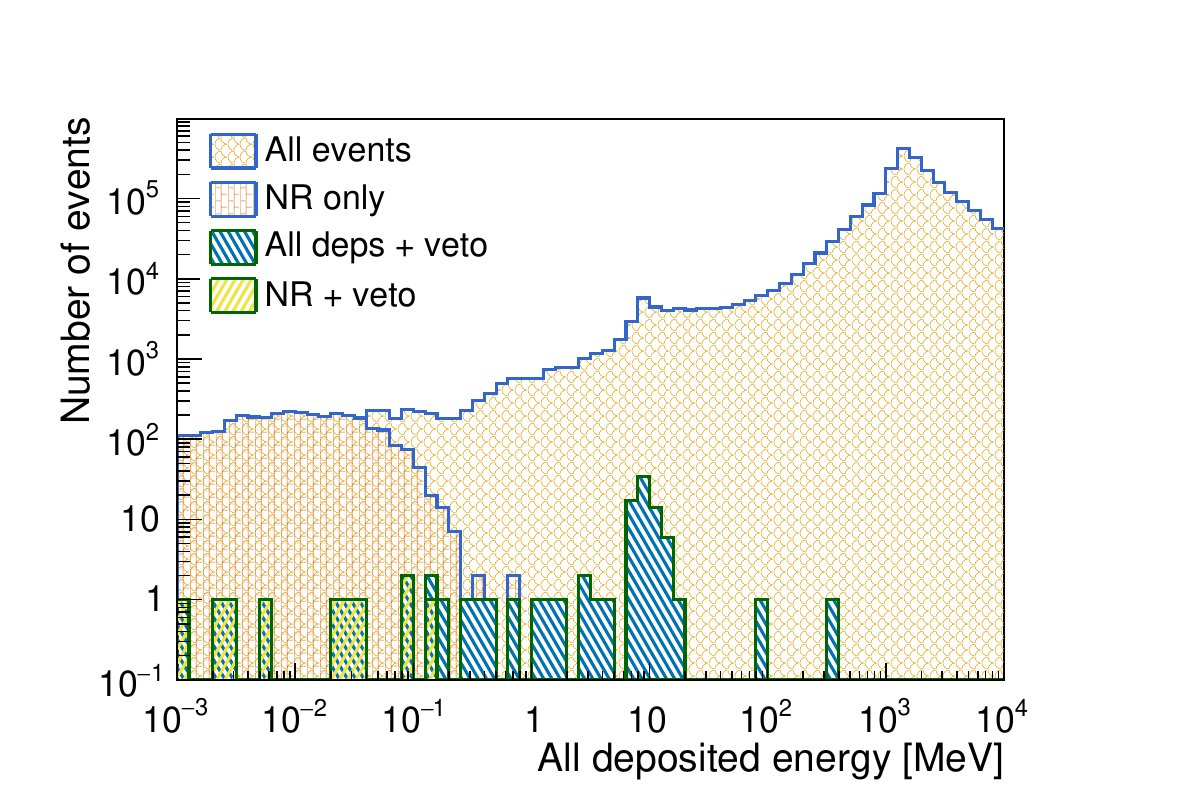}
    \caption{Energy depositions inside the TPC for simulations with polyhalite as the surrounding rock material. The distribution from all events with NR depositions is compared to the distribution from events with only NR. Added are also the same distributions but only for events passing the veto. Note the histograms are binned in logarithmic scale with 10~bins per decade in energy.}
    \label{fig:tpc_all_deps_polyhalite}
\end{figure}

Figure~\ref{fig:tpc_all_deps_polyhalite} shows the spectra of total energy depositions summed over all deposition types in the TPC for the simulation in the polyhalite rock. Distributions for all events and for events with only nuclear recoils are compared. Histograms for events after the veto cuts are also included. Almost all events below 50\,keV are coming from NRs before the veto cut is applied and all events below 100\,keV are NRs after the veto cut. Hence, considering only NR depositions in further analysis is justified.

\begin{figure}[!htb]
    \centering
    \makebox[10cm]{%
    \begin{subfigure}{0.49\textwidth}
        \centering
        \includegraphics[height=6cm]{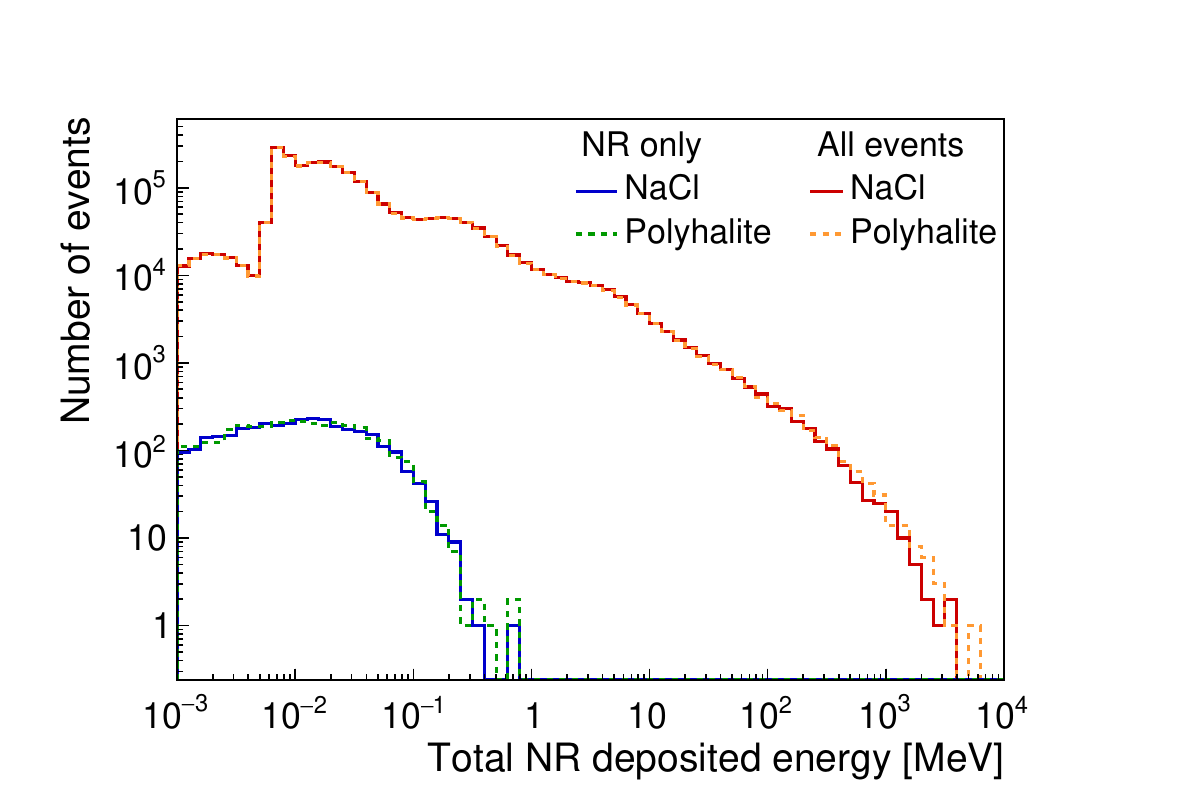}%
        \caption{}\label{fig:spectra-comparison:material}
    \end{subfigure}
    \begin{subfigure}{0.49\textwidth}
        \centering
        \includegraphics[height=6cm]{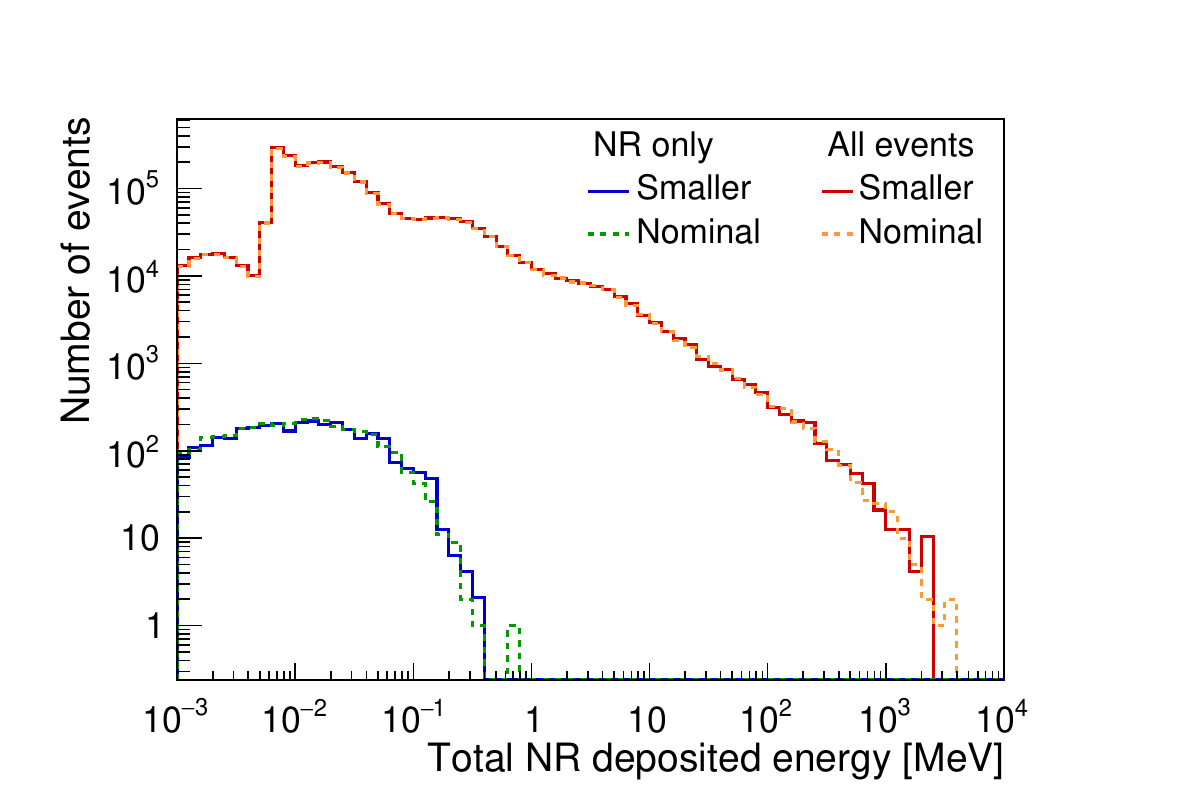}%
        \caption{}\label{fig:spectra-comparison:geom}
    \end{subfigure}
    }
    \caption{Spectra of energy depositions from nuclear recoils (NR) inside the TPC for the two types of rock (NaCl and polyhalite) and two cavern sizes. The red and orange histograms represent all simulated events in the sample that contain energy depositions from nuclear recoils. The blue and green histograms show all events where energy depositions other than from nuclear recoils were below the imposed threshold. \protect\subref{fig:spectra-comparison:material} Results of simulation with NaCl (solid line) and polyhalite (dashed line) as the rock material. \protect\subref{fig:spectra-comparison:geom} Results of simulation with the nominal (dashed line) and reduced (solid line) cavern sizes with NaCl chosen as the rock material. The sample with the smaller geometry was scaled up to the same equivalent exposure as for the nominal geometry.}
    \label{fig:spectra-comparison}
\end{figure}

Simulations for the two different types of rock for the two cavern sizes are compared in Fig.~\ref{fig:spectra-comparison:material} and Fig.~\ref{fig:spectra-comparison:geom}, respectively. 
Material composition definitely affects the neutron production (see Fig.~\ref{fig:n_energy:ch2}). However, we expect only the high-energy neutrons produced in the rock to reach the TPC. For these, the rock composition is not critical. This is confirmed by the similar shape in the distributions of energy depositions and absolute rate of events with only nuclear recoils in the TPC for the two rock compositions (after appropriate scaling to the same simulated exposure), demonstrated in Fig.~\ref{fig:spectra-comparison:material}.

Na\"{i}vely, the size of the cavern should not affect the neutron background for uniform and isotropic neutron emission. However, fast neutron emission (and we are concerned primarily with high-energy neutrons) is anisotropic \cite{wang2001} and simple considerations from diffusion theory may not apply. Moreover, neutron back-scattering  at the cavern walls \cite{carson2004}, which is important mostly for thermal and low-energy neutrons, may change the neutron distribution for caverns of different sizes. No noticeable difference in spectral shapes or absolute numbers were found for the nuclear recoil spectra for all events and for events with nuclear recoils only (after appropriate scaling to the same simulated exposure, see Fig.~\ref{fig:spectra-comparison:geom}).

It was found that only a very small number of produced neutrons reach the TPC volume without having any correlated signal in the veto systems (LXe skin, liquid scintillator or water tank). A small fraction of those originated directly in the primary muon interactions with the surrounding rock or with parts of the detector, and they produce signals in the active TPC volume within about \SI{1}{\ms} after the initial muon. A larger fraction of neutrons come from the activation of $^{17}$N from $^{19}$F within the PTFE-made field cage.
The activation process is similar to the production of $^{9}$Li and $^{8}$He in scintillators as reported by the KamLAND \cite{kamland2010}, Borexino \cite{borexino2013} and Daya Bay \cite{dayabay} collaborations. In our case, the neutron emission from $^{9}$Li and $^{8}$He in the scintillator is easily tagged by the detection of the electron from the beta decay. Also, the scintillator is further away from the TPC than the PTFE field cage and the lifetimes of $^{9}$Li and $^{8}$He are less than one second, making rejection of these events easier by requiring a delayed anti-coincidence with a muon. Neutrons from $^{17}$N decays (\SI{4.2}{s} lifetime) produce signals with a significant delay after the direct activity induced by the primary muon, and therefore they avoid any efficient veto from the observed muon. The simulation produces about 40 (1.3) delayed neutrons per \SI{1}{t} of PTFE per \SI{10}{years} at the location in NaCl (in polyhalite). The simulation considered approximately \SI{3.3}{\tonne} of PTFE in the field cage around the active region of the TPC. Smaller amounts of the material may help in reducing the background rate from this process.

The resulting numbers of selected neutron events are summarised in Table~\ref{tab:results}. After the full selection process described in the previous subsection (including multiple-scatter and fiducial-volume cut as in Table~\ref{tab:cuts}, lower part), no events passed in the sample with NaCl as the material of the surrounding rock. A single event passed the selection in the sample with polyhalite.
In the case where no liquid scintillator is used as an additional veto system, no events were observed for the site in salt, and a total of 2 events were observed for the site in polyhalite.
The table includes the estimated confidence intervals for event rates, based on the statistical uncertainties only.
The estimated rates in both cases are well below the expected physics background (from atmospheric neutrinos and two-neutrino double-beta decay) of a few tens of events in 10\,years extrapolated from the estimates in Table~6 of Ref.~\cite{LZexp2019}. 

\begin{table}[!htb]
    \caption{Results of MC simulations for the two considered locations. The column `Preselection' includes all events with nuclear recoils only (for the exposure time given in the 2nd column) before removing multi-scatter events and those which are outside of the fiducial volume. The column `Observed events' includes only single scatters in the fiducial volume. The upper part of the table refers to the analysis with the liquid scintillator veto. The lower part refers to the no-liquid-scintillator case. \textit{Unified} confidence intervals, as suggested in \cite{Feldman_1998}, are given for the number of background events in 10 years at 90\% CL and include only statistical uncertainty; systematic uncertainties in the muon flux and in the  neutron production yield were not included in this table.}
    \label{tab:results}
    \centering
    \small
    \begin{tabular}{@{}lccccr@{--}l@{}}
      \toprule
                              & Equivalent      & Preselection & Observed & Rate           & \multicolumn{2}{c}{}            \\
      Depth/Material          & exposure [year] &              & events   & [per 10 years] & \multicolumn{2}{c}{90\%\,CL}    \\
      \midrule
                              & \multicolumn{6}{c}{With liquid scintillator veto}                                            \\
      \cmidrule{2-7}
      2850~\mwe/NaCl       & 29              & 5            & 0        &                & \multicolumn{2}{c}{\num{<0.84}} \\
      3575~\mwe/polyhalite & 97              & 10           & 1        & 0.10           & 0.01 & 0.45                     \\
      \midrule
                              & \multicolumn{6}{c}{Without liquid scintillator veto}                                         \\
      \cmidrule{2-7}
      2850~\mwe/NaCl       & 29              & 27           & 0        &                & \multicolumn{2}{c}{$<$0.84}     \\
      3575~\mwe/polyhalite & 97              & 38           & 2        & 0.21           & 0.05 & 0.61                     \\
      \bottomrule
    \end{tabular}
\end{table}

Conservative systematic uncertainties due to neutron production are about a factor of 2. Uncertainties linked to the muon flux are about 10\% for the existing site where the flux has been measured (but may still be slightly different depending on the exact location of the laboratory) and about 20\% for a deeper site where the flux has been calculated based on the geophysical model of the Boulby mine but the exact location is not determined. We calculated the mean muon energies to be 259\,GeV and 282\,GeV for the 2850\,\mwe\ and 3575\,\mwe\ sites, respectively. Our simulations did not take this difference in the muon spectra into account and there is a small increase in the neutron production yield of (6--7)\% associated with such increase in the mean muon energy. This change is small compared with the other systematic uncertainties mentioned.

\begin{figure}[!htb]
    \centering
    \begin{subfigure}{0.32\textwidth}
    \centering
        \includegraphics[width=\textwidth]{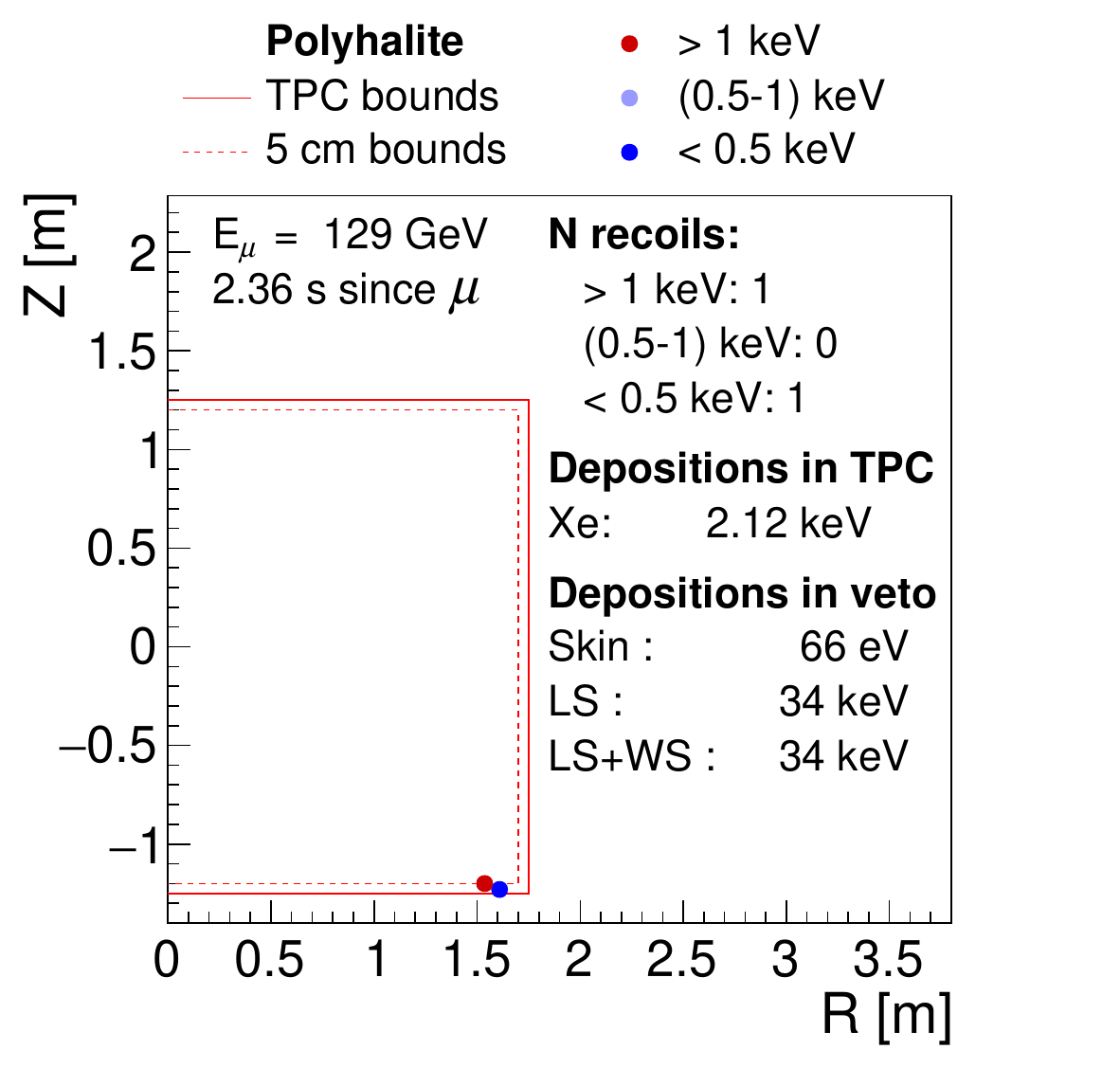}
        \caption{}\label{fig:evd:a}
    \end{subfigure}%
    \begin{subfigure}{0.32\textwidth}
        \centering
        \includegraphics[width=\textwidth]{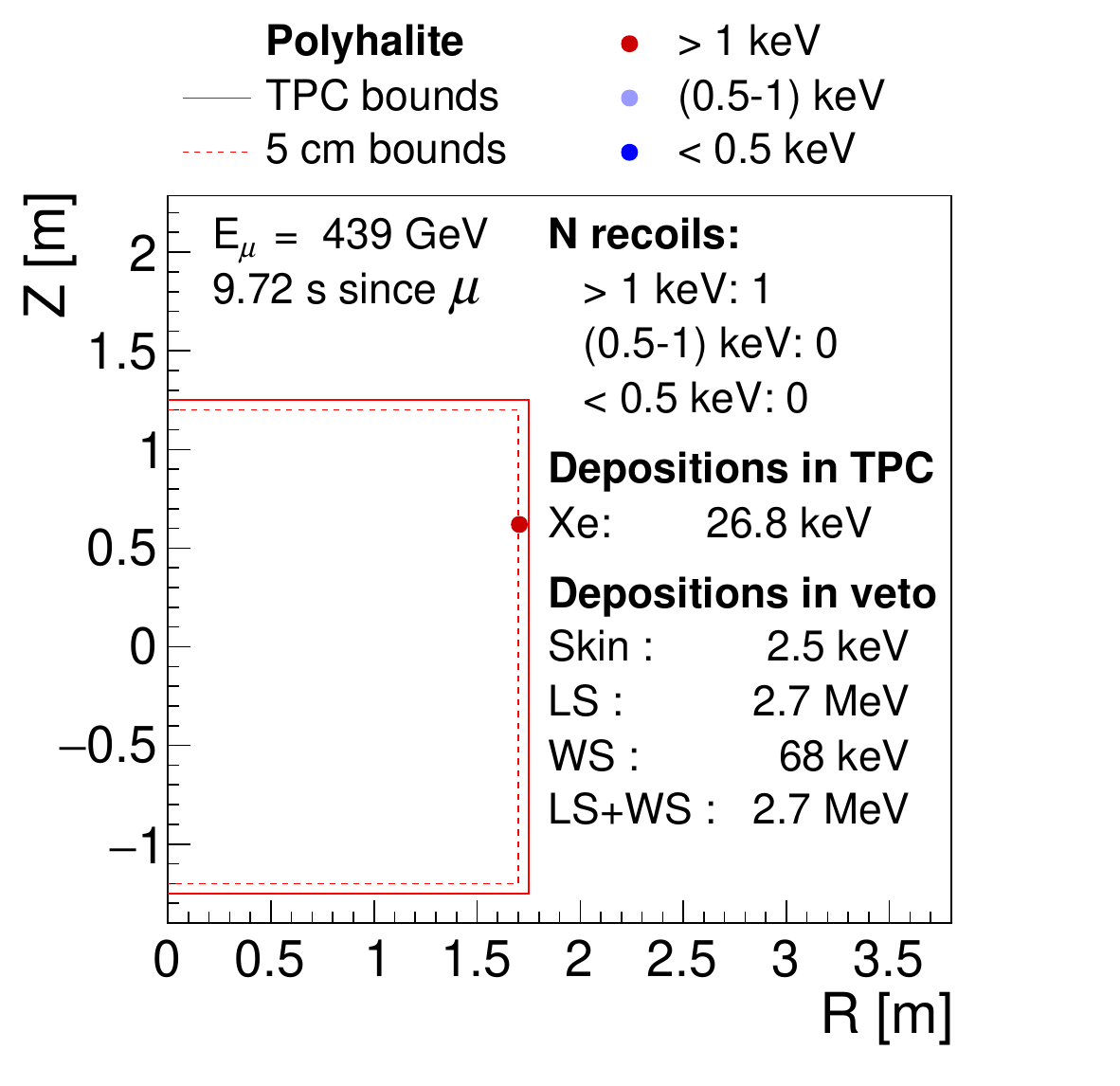}
        \caption{}\label{fig:evd:b}
    \end{subfigure}%
    
    \begin{subfigure}{0.32\textwidth}
        \centering
        \includegraphics[width=\textwidth]{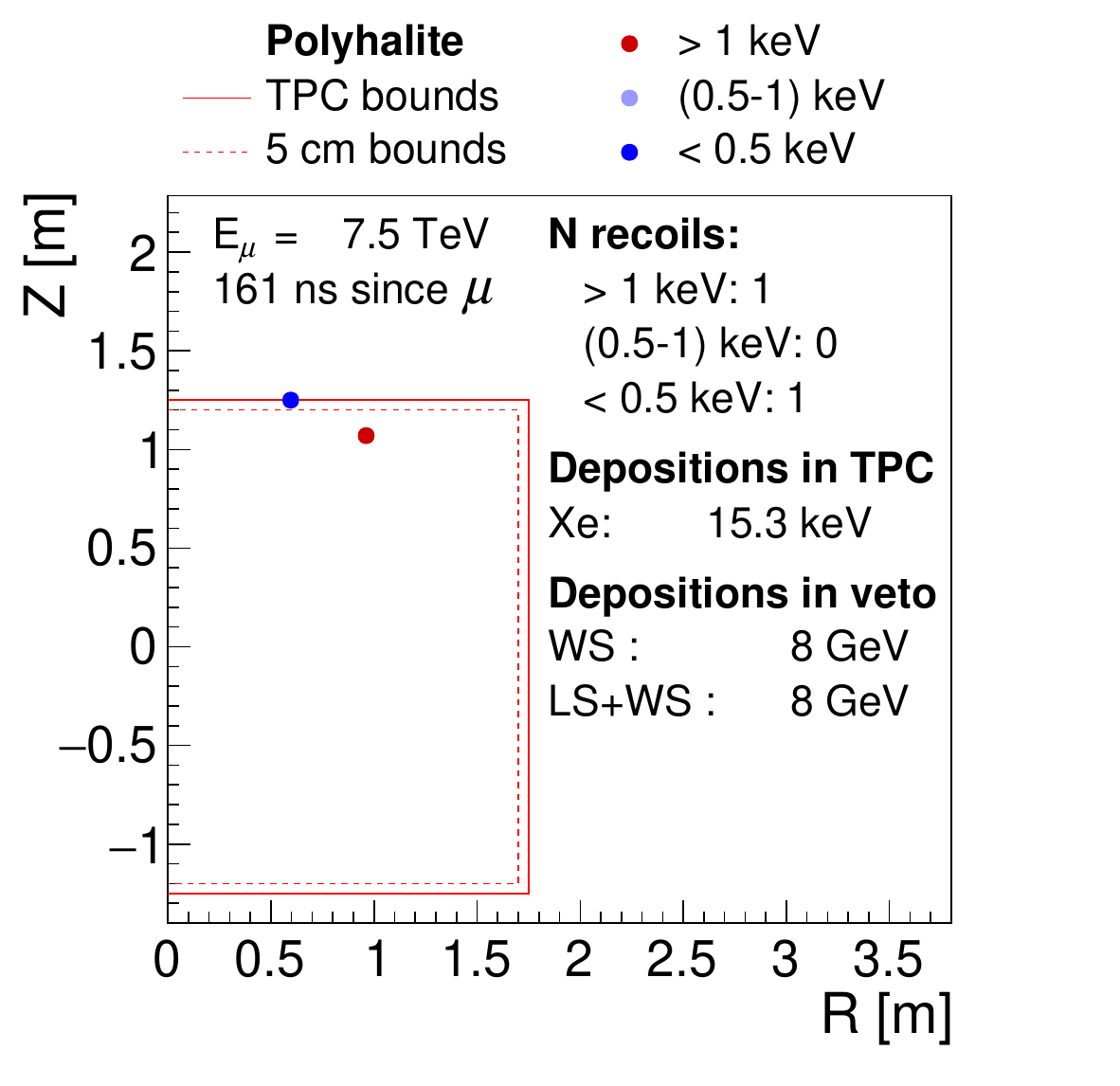}
        \caption{}\label{fig:evd:c}    
    \end{subfigure}
    \begin{subfigure}{0.32\textwidth}
        \centering
        \includegraphics[width=\textwidth]{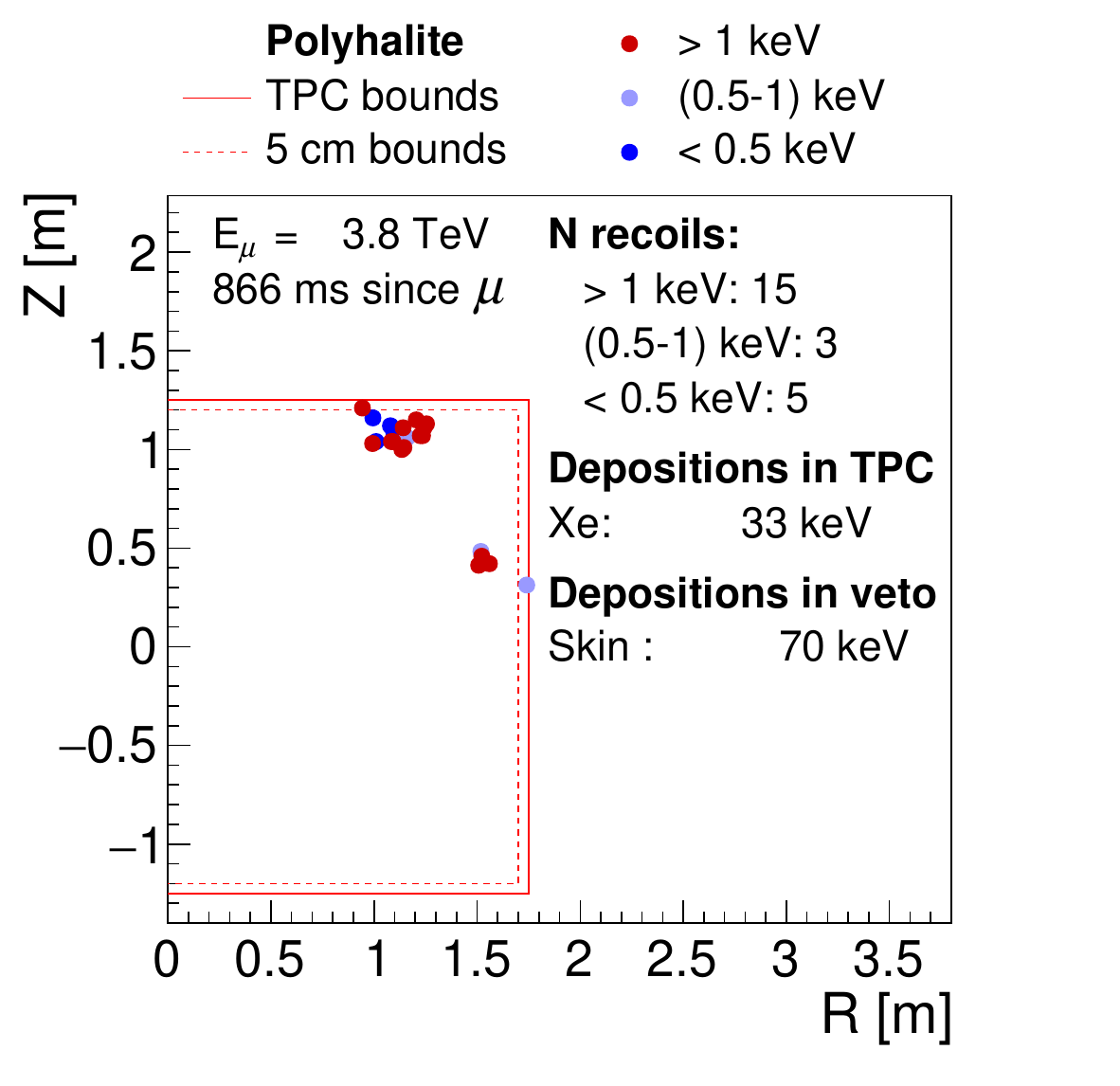}
        \caption{}\label{fig:evd:d}    
    \end{subfigure}

    \caption{Visualisations of individual nuclear recoils in some events of interest. Two events passing all signal selection criteria for simulations in polyhalite and within the scenario of no LS veto present are shown in \subref{fig:evd:a} and \subref{fig:evd:b}. The first event, \subref{fig:evd:a}, also passed the selection criteria within the scenario where the LS veto was considered. The event shown in \subref{fig:evd:c} was rejected based on the large coincident deposition in the WS and event \subref{fig:evd:d} was rejected based on the presence of multiple nuclear recoils above the assumed energy threshold of 0.5\,keV. Locations of individual recoils of Xe nuclei withing the active volume of the TPC are indicated in coloured markers. The vertical coordinate Z and radius R from the TPC's vertical axis are used. The colours indicate whether the recoil deposited energy of more than 1\,keV (red), between 0.5\,keV and 1\,keV (light blue), or below 0.5\,keV (blue). Each event visualisation includes additional information: initial energy of the simulated muon $E_\mu$, time of the energy depositions since the generation of the primary muon, number of recoils within the ranges of energy depositions described above, amounts of energy deposited by the nuclear recoils in the TPC (Xe), and amounts of energy deposited in the veto systems (LXe skin, liquid scintillator and water tank) if non-zero.}
    \label{fig:evd}
\end{figure}

A simplified visualisation of example events is shown in Fig.~\ref{fig:evd}. The one event which passed all the selection criteria, including the LS veto, for the sample in polyhalite is shown in Fig.~\ref{fig:evd:a}. The observed nuclear recoils are located at the boundary of the fiducial volume and are caused by a delayed neutron from the $^{17}$N activation in PTFE. The activation in a muon-induced hadronic shower was a result of $\pi^-$ absorption on $^{19}$F. The small coincident depositions in the veto systems never crossed the required threshold. No other delayed activity was recorded within the TPC. The event passing our selection for the case with no LS veto is shown in Fig.~\ref{fig:evd:b}. The single nuclear recoil is at the boundary of the fiducial volume. The coincident energy deposition in the LS volume caused it to be rejected in the scenario with the LS veto, however, the 2.7\,MeV deposition was insufficient to trigger a WT veto. Similarly to the former event, the recoil was caused by a delayed neutron from the activation in the PTFE. The activation was due to a neutron from a muon-induced hadronic shower within the PTFE. An example event which was rejected due to the veto from the WT is shown in Fig.~\ref{fig:evd:c}. The nuclear recoils within the detector were initiated by a neutron originating from a muon-induced hadronic shower in the polyhalite rock. Figure~\ref{fig:evd:d} shows an example of an event which was rejected due to the presence of multiple nuclear recoils within the TPC.

%

\section{Conclusions}\label{conlusions}
%

The goal of the work presented here was to investigate the implication of laboratory depth on the muon-induced background in a future dark matter, xenon-based experiment capable of reaching the so-called neutrino floor. As a case study, we considered two locations at the Boulby Underground Laboratory (UK): an experimental cavern in salt at a depth of 2850\,\mwe, and a deeper laboratory located in polyhalite rock at a depth of 3575\,\mwee. These depths are similar to other underground laboratories around the world, and our conclusions apply to those with straightforward scaling for the actual muon flux. We have carried out detailed simulations of cosmogenic background in a simplified experimental geometry with the \geant{} simulation toolkit.

We have tested muon-induced neutron production in \geant{} version 10.5 and compared the results to previous versions and available measurements. This allowed us to evaluate a conservative systematic uncertainty of our simulations to be about a factor of 2.

We have performed simulations for an experiment similar in configuration to an scaled-up LZ detector. The detector model contained $\sim$100\,tonnes of LXe with 70\,tonnes of active mass, surrounded by a LXe `skin' and an additional veto system. We conclude that, after applying a standard simplified analysis procedure and cuts, the event rate caused by cosmogenic activity stays below 1 event per 10\,years in the fiducial volume of the LXe-TPC (64\,tonnes). This rate is well below the expected background of tens of events from ERs/NRs from physics backgrounds such as two-neutrino double beta decay of $^{136}$Xe and solar/atmospheric neutrinos with ER events leaking into NR band due to limited discrimination. From the point of view of cosmogenic background, a depth of about 3\,k\mwe\ or deeper is sufficient for a next-generation dark matter experiment based on liquid xenon. The observed residual background of NR events comes from the production and delayed $\beta-n$ decay of $^{17}$N in PTFE (on fluorine) where only neutron scattering is detected. Our material budget contained about \SI{2.8}{\tonne} of PTFE. Although the residual background is very low, the design of a future experiment may need to limit PTFE usage to the necessary minimum. 

We have also investigated two veto system configurations: a default one with instrumented liquid scintillator surrounding the cryostat, and an option without the scintillator. No significant difference was observed between the two scenarios, which lead us to conclude that the additional veto system is not required to suppress cosmogenic backgrounds for the goal sensitivity at the studied depth. This conclusion, however, does not apply to other types of backgrounds where liquid scintillator is particularly efficient in tagging neutron events from detector components. 

\section*{Acknowledgements}
We acknowledge funding from the UKRI Science \& Technology Facilities Council under the 2019 Opportunities Call to conduct this study (ST/T002786/1, ST/T002107/1).

For the purpose of open access, the authors have applied a Creative Commons Attribution (CC BY 4.0) licence to any Author Accepted Manuscript version arising from this submission.

\printbibliography

\end{document}